\documentclass{emulateapj}
\usepackage{apjfonts}

\newcommand{\chandra}{{\it Chandra}}
\def\gs{\mathrel{\raise1.16pt\hbox{$>$}\kern-7.0pt %
\lower3.06pt\hbox{{$\scriptstyle \sim$}}}}         %
\def\ls{\mathrel{\raise1.16pt\hbox{$<$}\kern-7.0pt %
\lower3.06pt\hbox{{$\scriptstyle \sim$}}}}         %

\slugcomment{}
\shorttitle{Optical IDs for the XBootes Survey}
\shortauthors{}

\begin{document}

\title{The {\it Chandra} XBo\"otes Survey - III: Optical and Near-IR Counterparts}

\author{Kate Brand\altaffilmark{1}, Michael J. I. Brown\altaffilmark{2}, Arjun Dey\altaffilmark{1}, Buell T. Jannuzi\altaffilmark{1}, Christopher S. Kochanek\altaffilmark{3}, Almus T. Kenter\altaffilmark{4}, Daniel Fabricant\altaffilmark{4}, Giovanni G. Fazio\altaffilmark{4}, William R. Forman\altaffilmark{4}, Paul J. Green\altaffilmark{4}, Christine J. Jones\altaffilmark{4},  Brian R. McNamara\altaffilmark{3}, Stephen S. Murray\altaffilmark{4}, Joan R. Najita\altaffilmark{1}, Marcia Rieke\altaffilmark{5}, Joseph C. Shields\altaffilmark{6}, Alexey Vikhlinin\altaffilmark{4}}

\altaffiltext{1}{National Optical Astronomy Observatory, Tucson, AZ 85726-6732; brand@noao.edu} 
\altaffiltext{2}{Princeton University Observatory, Peyton Hall, Princeton, NJ 08544}
\altaffiltext{3}{Department of Astronomy, The Ohio State University, 140 West 18th Avenue, Columbus, OH 43210}
\altaffiltext{4}{Harvard-Smithsonian Center for Astrophysics, 60 Garden Street, Cambridge, MA 02138}
\altaffiltext{5}{Steward Observatory, University of Arizona, 933 North Cherry Avenue, Tucson, AZ 85721}
\altaffiltext{6}{Department of Physics and Astronomy, Ohio University, Athens, OH 45701}

\begin{abstract}

The XBo\"otes Survey is a 5-ks \chandra\ survey of the Bo\"otes Field of the NOAO Deep Wide-Field Survey (NDWFS). This survey is unique in that it is the largest (9.3 deg$^2$), contiguous region imaged in X-ray with complementary deep optical and near-IR observations. We present a catalog of the optical counterparts to the 3,213 X-ray point sources detected in the XBo\"otes survey. Using a Bayesian identification scheme, we successfully identified optical counterparts for 98\% of the X-ray point sources. The optical colors suggest that the optically detected galaxies are a combination of $z<$1 massive early-type galaxies and bluer star-forming galaxies whose optical AGN emission is faint or obscured, whereas the majority of the optically detected point sources are likely quasars over a large redshift range. Our large area, X-ray bright, optically deep survey enables us to select a large sub-sample of sources (773) with high X-ray to optical flux ratios (f$_x$/f$_o>$10). These objects are likely high redshift and/or dust obscured AGN. These sources have generally harder X-ray spectra than sources with 0.1$<$f$_x$/f$_o<$10. Of the 73 X-ray sources with no optical counterpart in the NDWFS catalog, 47 are truly optically blank down to $R\sim$25.5 (the average 50\% completeness limit of the NDWFS $R$-band catalogs). These sources are also likely to be high redshift and/or dust obscured AGN. 

\end{abstract}


\section{Introduction}

Active Galactic Nuclei (AGN) are complex objects which radiate across the entire electromagnetic spectrum. To gain a better insight into their nature and how they evolve, we require large samples of AGN in multiple wave-bands. Previous studies at optical and soft X-ray wavelengths have failed to obtain a complete census of the AGN population because dust and gas obscures many AGN from view. In fact in the local Universe, optically obscured AGN may outnumber optically unobscured AGN by a factor of four \citep{mai95}. In all but the shallowest surveys, the number density of AGN identified in the hard X-ray and mid-IR bands is far greater than is found in optical surveys of comparable depth (\citealt{bau04}; \citealt{ris04}; \citealt{ste05}). Thus it is important to determine the extent to which there exists a hidden population of AGN whose optical properties do not identify them as AGN. Because of the ability of hard X-rays to penetrate dust and all but the most extreme column densities of gas, X-ray surveys with sensitivities above $\sim$4 keV can provide relatively unbiased samples of AGN over a range of redshifts. 

We summarise some of the existing extragalactic X-ray surveys in Table~\ref{tab:xsurveys}. The \chandra\ \citep{wei02} and {\it XMM-Newton} \citep{jan01} observatories have led to great advances in X-ray astronomy, producing surveys which are 10 - 100 times deeper than those by previous X-ray telescopes. \chandra 's imaging resolution is superior to all previous and current X-ray telescopes (\citealt{van97}; \citealt{gar03}); this is crucial for determining the correct optical counterpart in deep optical imaging data in which the surface density of sources is large. While the deepest surveys have resolved $\sim$80\% of the hard (2-7 keV) X-ray background into discrete sources (\citealt{bra01}; \citealt{ros02}; \citealt{mor03}; \citealt{wor04}) and an even larger fraction of the soft (0.5-2 keV) X-ray background, they typically cover only small ($\sim$0.1 deg$^2$) areas of the sky. Surveys covering a larger volume are needed to overcome cosmic variance and to better determine the properties of the most luminous and rarest sources such as powerful quasars, whose number densities are low. Large contiguous areas and extensive multi-wavelength coverage are also necessary for detailed studies of AGN clustering and environment.

\begin{deluxetable*}{lllll}
\tabletypesize{\scriptsize}
\setlength{\tabcolsep}{0.02in}
\tablecolumns{4} 
\tablewidth{0pc} 
\tablecaption{\label{tab:xsurveys} Properties of recent extragalactic deep and medium-deep X-ray surveys ordered by increasing survey area. The X-ray flux is quoted for the 0.5-2 keV band and has been converted to the flux assuming a power-law with $\Gamma$=1.7 where needed. The optical magnitude limit is taken from the referenced paper and may be defined in different ways. {\bf a.} The LALA field is in the north-east corner of the Bo\"otes field. {\bf b.} The number in parenthesis indictates the total expected area of the survey when completed.} 
\tablehead{ 
\colhead{Survey} & \colhead{Area} & \colhead{X-ray Flux limit} & \colhead{$R$ limit} & \colhead{Reference} \\
\colhead{} & \colhead{deg$^2$} & \colhead{${\rm ergs~s^{-1}~cm^{-2}}$} & \colhead{magnitudes} & \colhead{}
}
\startdata  
Chandra Deep Field North & 0.1 & 2.4$\times 10^{-17}$ & $\sim$26.4 &\citet{bar03}, \citet{ale03}\\
Chandra Deep Field South & 0.1 & 5.7$\times 10^{-17}$ & $\sim$26.5 &\citet{gia02}\\
Extended Chandra Deep Field South & 0.3 & 8$\times 10^{-17}$ &$\sim$26.5 & \citet{leh05}\\
Chandra Extended Growth Strip (DEEP)& 0.1 & 1.0$\times 10^{-16}$ & 23.4 & \citet{nan05}\\
Chandra LALA$^{\bf a}$ & 0.1 & 1.4$\times 10^{-16}$ & 25.7 & \citet{wan04}\\
SPICES-II & 0.1 & 1.8$\times 10^{-16}$ &25.4 & \citet{ste02}\\
XMM Lockman hole & 0.25 & 3.0$\times 10^{-16}$ & $\sim$27.1 &\citet{has01},\citet{wil04}\\
CLASXS & 0.36 & 5.0$\times 10^{-16}$ & 27.0 & \citet{yan04}\\
Chandra survey of 13h XMM/ROSAT  &0.25 & 5.0$\times 10^{-16}$& 27.0 &\citet{mch03}\\
Chandra Multi-wavelength Project & 0.8 (14.0$^{\bf b}$) & variable & 24.0 & \citet{kim04}, \citet{gre04}\\
ROSAT Ultra Deep & 0.36 & 1.1$\times 10^{-15}$ & 25.5 & \citet{leh01}\\
{\bf XBo\"otes} & {\bf 9.3} & ${\bf 4.0\times 10^{-15}}$ & {\bf 25.5 } & Murray et al.~(2005), Kenter et al.~(2005); this work\\
XMM-Newton/2dF & 1.5 & 4.0$\times 10^{-15}$& 19.5 & \citet{geo03}\\
HELLAS2XMM & 0.9 (4.0$^{\bf b}$)&7.5$\times 10^{-15}$& 24.5 & \citet{bal02}\\
\hline
\enddata
\end{deluxetable*}

Here we identify the candidate optical counterparts to the X-ray point sources in a new, medium depth (5-ks/pointing), wide-field X-ray survey performed with \chandra\  ACIS-I (XBo\"otes; Murray et al. 2005) of the Bo\"otes field of the NOAO Deep Wide-Field Survey (NDWFS; \citealt{jan99}; Jannuzi et al. in prep.; Dey et al. in prep.). The XBo\"otes survey is unique because of its contiguous imaging coverage over a large (9.3 deg$^2$) area with {\it Chandra}. While not as deep as many past surveys made with {\it Chandra} and {\it XMM}, the large co-moving volume surveyed allows us to obtain large samples of sources comprising the bright end of the X-ray luminosity function and to determine the nature of rare populations whose number densities are too small to obtain meaningful statistics in small area surveys. The large contiguous area is also critical for studies of AGN clustering: one of the key goals of the XBo\"otes survey. We can also perform stacking analyzes of the X-ray data to determine the mean X-ray properties of different populations selected in different wavebands (e.g., \citealt{bra05}; Watson et al.~in prep.). The Bo\"otes field is unique in the extent of its multi-wavelength coverage over such a large area. The multi-wavelength data includes X-ray (\chandra), UV ({\it GALEX}; \citealt{hoo04}), optical (NDWFS), near-IR (FLAMEX; \citealt{els05}), mid-IR ({\it Spitzer}/IRAC --  \citealt{eis04}; {\it Spitzer}/MIPS --  \citealt{soi04}), and radio (VLA/FIRST; \citealt{bec95} and WSRT; \citealt{dev02}). Optical spectroscopic follow-up observations have also been undertaken for all X-ray sources with $I<$21.5 as part of the AGES survey (Kochanek et al. in prep.). 

In Murray et al.~(2005; Paper I), we described the general characteristics of the X-ray survey and determined the angular clustering of the sources. In Kenter et al.~(2005; Paper II) we presented the X-ray catalog. The main aim of this paper is to provide an accurate catalog of the optical counterparts of the X-ray point sources in the Bo\"otes region which were presented in Kenter et al.~(2005). We begin with a summary of the X-ray, optical, and near-IR data used in the XBo\"otes and NDWFS surveys (Section~2). We discuss our method of associating the X-ray source with optical counterparts in Section~3. The robustness of the resulting matched catalog as well as the properties of the catalog are described in Section~4. In Section~5, we describe the X-ray and optical properties of the matched catalog. The main points of the paper are summarized in Section~6.  

\section{The Data}

\subsection{X-ray Imaging: the XBo\"otes Survey}

The details of the X-ray observations for the XBo\"otes Survey and the resulting XBo\"otes source catalog have been presented in Murray et al.~(2005) and Kenter et al.~(2005). In this Section, we summarize the main survey characteristics.

A 9.3 deg$^2$ region of sky chosen to match the area covered with NDWFS was observed by the Advanced CCD Imaging Spectrometer (ACIS-I) on the {\it Chandra X-ray Observatory} over a 2 week time interval in March and April 2003. The data was taken in 126 separate pointings, each observed for $\approx$ 5-ks. The CIAO~3.0.2 wavelet detect algorithm ({\small WAVDETECT}; \citealt{fre02}) was used to detect X-ray sources in the total (0.5-7.0 keV) band data. A probability threshold of 5$\times 10^{-5}$ was chosen as the best compromise between maximizing the completeness while minimizing the number of spurious detections. Although the {\small WAVDETECT} algorithm provides a robust method of identifying real X-ray sources, it performs less well at providing an accurate source position and its uncertainties. The positions were instead estimated by iteratively centroiding the counts within the 50\% encircled-energy radius of the \chandra\ ACIS point spread function. The total counts from each X-ray source were determined within the 90\% encircled energy radius (see Murray et al.~(2005) for details). No corrections were made for the remaining flux expected to fall outside this radius. The counts in the soft (0.5-2.0 keV) and hard (2.0-7.0 keV) bands were determined in the soft and hard-band images using the position and 90\% encircled energy radius from the total band image. The X-ray catalog comprises 3,293 unique X-ray sources with $\ge$4 counts in the total band images (Kenter et al.~2005). We expect only $\sim$35 of these sources to be spurious in the full survey (Kenter et al.~2005). For the matching with cataloged optical counterparts, we only considered the 3,213 X-ray sources which overlap with the NDWFS area (see Figure~\ref{fig:radec}).

To estimate the X-ray flux for each source, a local background was calculated and subtracted from the net counts. Telescope vignetting results in an effective exposure time that varies with X-ray energy and position on the image (dropping to $\sim$20\% near the edges of the field). To account for this, we multiply the background-subtracted counts by the ratio of the average exposure time (5-ks) to the effective exposure time at the source position. We also divide by the fractional effective area at the source position. Assuming a power-law spectrum with photon index $\Gamma$=1.7 and the Galactic HI column density of 1$\times$10$^{20}$ cm$^{-2}$ \citep{sta92}, a 4 count source in our survey corresponds to 7.8$\times$10$^{-15}$ ergs cm$^{-2}$ s$^{-1}$, 4.7$\times$10$^{-15}$ ergs cm$^{-2}$ s$^{-1}$, and 1.5$\times$10$^{-14}$ ergs cm$^{-2}$ s$^{-1}$ in the 0.5-7 keV, 0.5-2 keV and 2-7 keV bands respectively. This corresponds to a total band X-ray luminosity of L$_x= 2 \times 10^{41} {\rm ergs~s^{-1}}$, $4 \times 10^{43} {\rm ergs~s^{-1}}$, and $6 \times 10^{44} {\rm ergs~s^{-1}}$, at $z$=0.1, 1, and 3 respectively.


\subsection{Optical and Near Infra-red Imaging: the NDWFS}

The NOAO Deep Wide-Field Survey (NDWFS) comprises deep optical and near infra-red imaging in two $\sim$9 deg$^2$ regions of the sky in Bo\"otes and Cetus, and is designed to study the formation and evolution of large-scale structure (\citealt{jan99}; \citealt{bro03}). The optical and near-IR data for the Bo\"otes field were obtained using the 4m and 2.1m telescopes of the Kitt Peak National Observatory between February 1998 and April 2004. In this paper, we consider $B_W, R$, and $I$-band imaging over almost the entire 9.3 deg$^2$ of the NDWFS Bo\"otes field (the field covered by our X-ray imaging; Jannuzi et al. in prep.), and $K$-band imaging over approximately 6 deg$^2$ of the Bo\"otes field (Dey et al. in prep.). $K$-band imaging has been obtained for the entire Bootes field, but only
6 deg$^2$ were available at the time of this work. We use the NDWFS data release version 3.0 which can be obtained through the NOAO science archive (http://archive.noao.edu/nsa/) and NDWFS homepage (http://www.noao.edu/noao/noaodeep). We determined the 50\% completeness limits as a function of magnitude by adding artificial stellar objects to copies of the data and recovering them with SExtractor. Details of the completeness as a function of NDWFS sub-field can be found in Jannuzi et al. (in prep.) and at the NDWFS homepage.  The average 50\% completeness limits of the $B_W$, $R$, $I$, and $K$-bands images are 26.7, 25.5, 24.9, and 18.6 respectively. We use Vega magnitudes throughout. 

The optical catalogs were generated using SExtractor, version 2.3.2 \citep{ber96} run in single-image mode with the minimum detection area, convolution filter, and signal above sky threshold optimized to provide a deep yet reliable catalog. The astrometry for the NDWFS was calibrated using matches to the USNO-A2 catalog \citep{mon98}, resulting in an uncertainty in the absolute positions of the sources of $< 0\farcs1$ (Jannuzi et al. in prep., Dey et al. in prep.). Detections in two separate optical bands were matched if the centroid of a detection in one band fell within the image ellipse calculated from a detection in the other band or if the centroids of the detections were within $1^{\prime\prime}$ of each other. In the case of a detection in the $R$-band, we used this to assign the position of the matched catalog; otherwise we used the position from either the $B_W$ or $I$-band catalogs in that order of preference. We will generally use this ordering wherever we must assign a optical parameter to a source. Because the depth of the $K$-band data is not comparable to the optical data and may contain a higher number of spurious sources, we required there to be a match in at least one of the optical bands in addition to the $K$-band detection for a source to be included in the single matched catalog. 

\section{Matching the X-ray and Optical Surveys}
\label{sec:match}

\begin{figure*}
\begin{center}
\setlength{\unitlength}{1mm}
\begin{picture}(150,120)
\put(0,-20){\special
{psfile="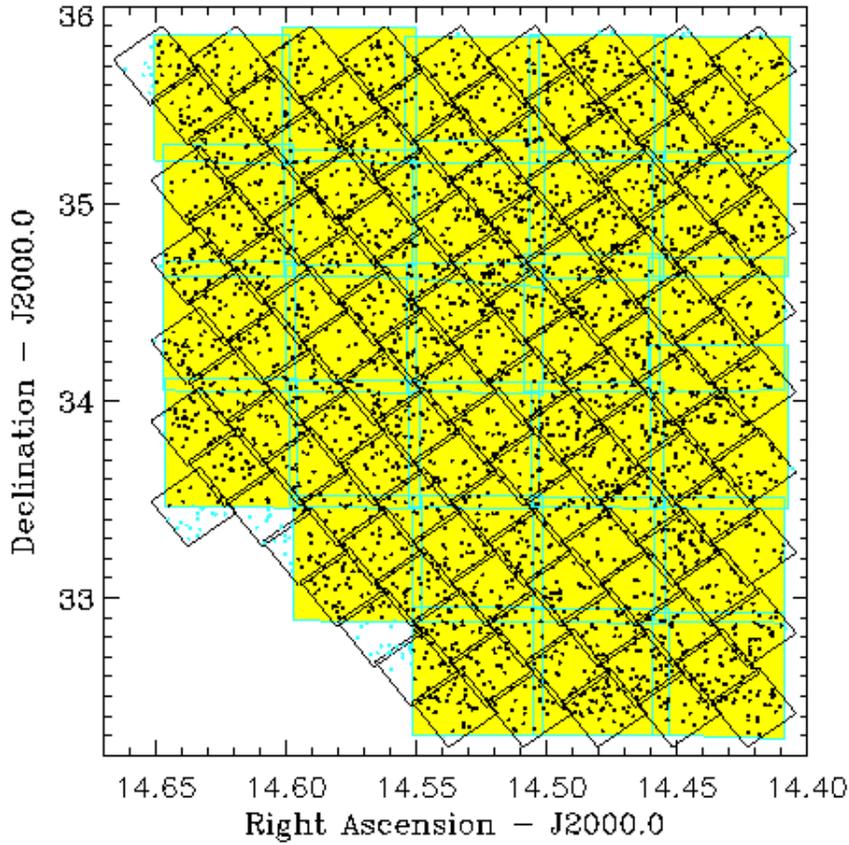" 
vscale=70 hscale=70 angle=0}}
\end{picture}
\end{center}
{\caption[junk]{\label{fig:radec} The spatial distribution of X-ray sources (denoted by black circles). Also plotted are the positions of the 126 X-ray pointings of the \chandra\ ACIS camera (diamond shaped boxes) and the 27 optical imaging sub-fields ($R$-band images; large shaded area). X-ray sources whose positions are outside the boundaries of the NDWFS Bo\"otes field are denoted with lightly shaded circles and are excluded from the analysis of this paper.}}
\end{figure*}

This paper is concerned with the optical counterparts of X-ray point sources in the NDWFS Bo\"otes field. It comprises 27 optical imaging sub-fields (each roughly corresponding to the area of a single pointing of the KPNO MOSAIC-1 prime-focus imager) and 126 pointings of the \chandra\ ACIS-I camera. The positions and overlap of the sub-fields on the sky can be seen in Figure~\ref{fig:radec}. Over-plotted are the 3,293 X-ray point sources with $\geq$ 4 counts. Although the surveys are extremely well matched in their spatial coverage, the difference in the size and shape of the individual optical and X-ray pointings led to small differences in their coverage. The X-ray sources represented by light dots are those that fall off the edge of the NDWFS survey; these sources are not included in any further analysis. The total number of X-ray sources considered in this paper is therefore 3,213.

\subsection{Relative Positional Offsets between X-ray and NDWFS Imaging Frames}

Any systematic offsets between the astrometric reference frames used for the X-ray and optical imaging or errors in the astrometric zero-point of any of the individual optical sub-fields or X-ray pointings could adversely affect our procedure for matching sources. The NDWFS images were placed on an astrometric frame defined by the fainter stars in the USNOA2.0 (The ACT astrometric frame; \citealt{mon98}; see Jannuzi et al. in prep. for details of NDWFS astrometry). The X-ray astrometry is based on the AGASC catalog \citep{sch03} and is referenced to the Tycho2 astronomic frame, which incorporates and supercedes the ACT astrometric frame. Although the X-ray and optical data should be on the same astrometric frame, we need to check and correct for any systematic offsets that may be present. Because of the large number of X-ray sources with cataloged optical counterparts (see Section~\ref{sec:opt_cparts}), we can use the data to determine and correct any systematic offset. 

We calculated the median offset between the positions of all 3213 X-ray sources and the positions of all matched optical sources in the survey. For this process, we used a simple scheme in which we take the nearest optical source within either 1.5$^{\prime\prime}$ or the 50\% X-ray positional uncertainty to be the matched optical counterpart. The median offset for sources in all the X-ray pointings was 0$\farcs$03$\pm$0$\farcs$02 east and 0$\farcs$53$\pm$0$\farcs$02 north of the optical positions.  

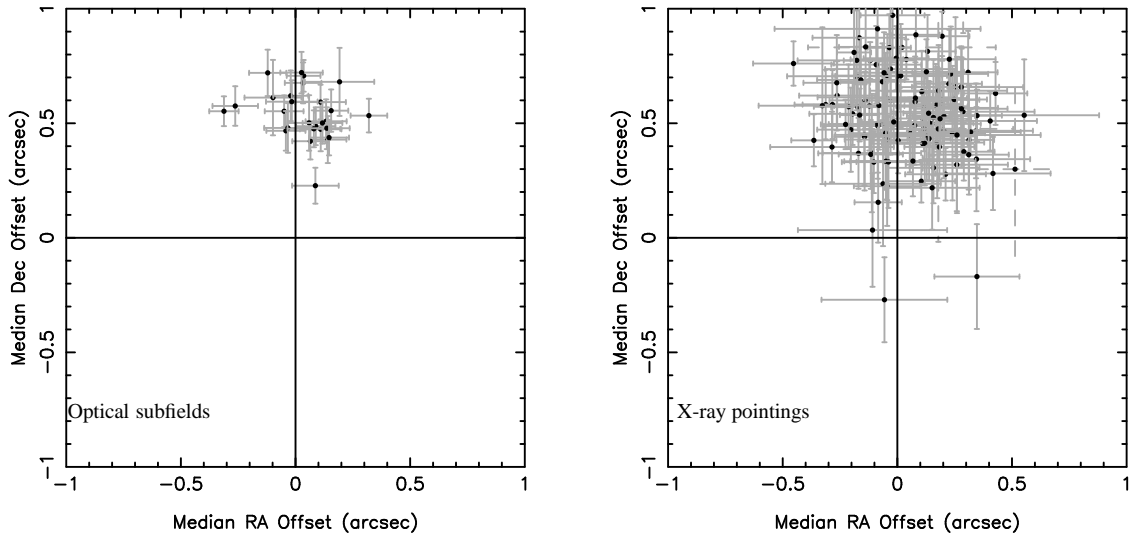
\begin{figure*}
\begin{center}
\setlength{\unitlength}{1mm}
\begin{picture}(100,70)
\put(-40,-15){\special
{psfile="f2a.ps" 
vscale=50 hscale=50 angle=0}}
\put(40,-15){\special
{psfile="f2b.ps" 
vscale=50 hscale=50 angle=0}}
\put(32,75){Before astrometry correction}
\put(-15,15){Optical subfields}
\put(66,15){X-ray pointings}
\end{picture}
\end{center}
{\caption[junk]{\label{fig:offset} The median X-ray to optical astrometric offsets and their corresponding errors for X-ray sources with optical counterparts in each of the 27 optical sub-fields (left) and each of the 126 X-ray pointings (right). The 5 X-ray pointings containing less than 15 matched sources are represented by dashed error bars. In these cases, we corrected the source positions using the median offset found for all the X-ray pointings.
}}
\end{figure*}

\begin{figure*}
\begin{center}
\setlength{\unitlength}{1mm}
\begin{picture}(100,80)
\put(-40,-15){\special
{psfile="f3a.ps" 
vscale=50 hscale=50 angle=0}}
\put(40,-15){\special
{psfile="f3b.ps" 
vscale=50 hscale=50 angle=0}}
\put(32,75){After astrometry correction}
\put(-15,15){Optical subfields}
\put(66,15){X-ray pointings}

\end{picture}
\end{center}
{\caption[junk]{\label{fig:offset_corr} The median X-ray to optical astrometric offsets and their corresponding errors for X-ray sources with optical counterparts in each of the 27 optical sub-fields (left) and each of the 126 X-ray pointings (right) after the astrometric correction was applied. 
}}
\end{figure*}
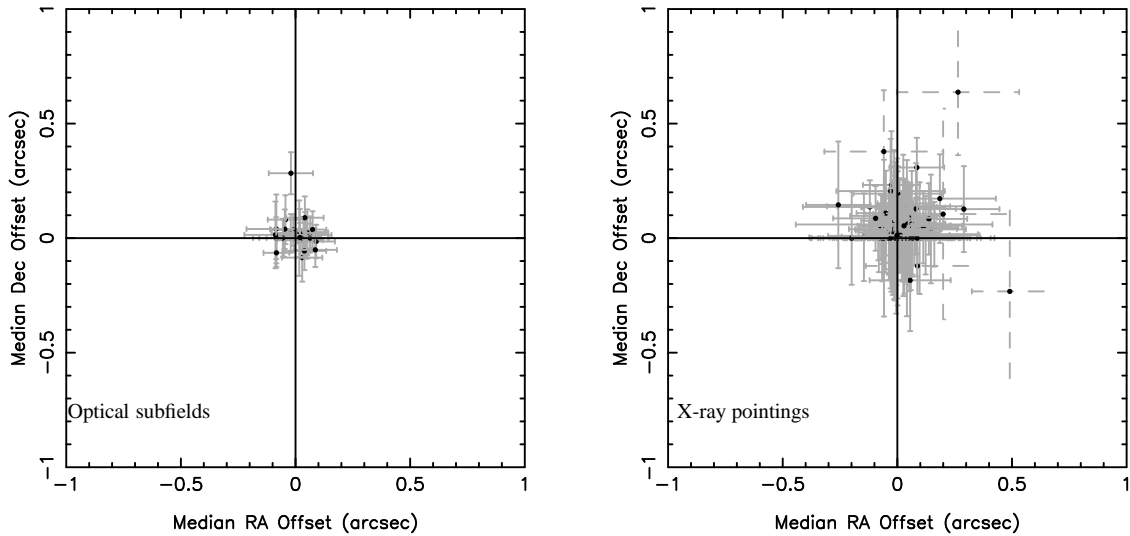

We determined whether there were any field-to-field variations in the astrometric zero-point by calculating the X-ray to optical positional offset for the sources detected within each X-ray pointing and for the sources detected within each optical sub-field. To identify this systematic offset we determined the median offset of all X-ray sources with a positional uncertainty of less than 1$^{\prime\prime}$ in a given \chandra/ACIS pointing. By correcting the positional offset of each X-ray pointing individually, we are correcting for any positional uncertainties between the individual pointings as well as any astrometric zero-point error between the X-ray and optical surveys. In Figure~\ref{fig:offset}, we show the median X-ray to optical astrometric offset and its corresponding error for X-ray sources with optical counterparts in each of the 27 optical sub-fields and in each of the 126 X-ray pointings. Because each optical sub-field contains approximately 4 X-ray pointings, the optical sub-fields will have astrometric offsets correlated with that of the X-ray pointings but will have smaller uncertainties due to the larger number of sources in each field. We therefore assumed no field-to-field variation in the astrometric offset of each optical sub-field and only corrected for any field-to-field variation in the astrometric offset of each X-ray pointing. In practice, we only adjusted the X-ray source positions of the sources if there were 15 or more X-ray sources with a positional uncertainty of less than 1$^{\prime\prime}$ (corresponding to sources with high X-ray counts and/or sources near the center of the X-ray pointing). In all pointings with less than 15 such sources, we used the median offset of sources in the full sample. The largest correction we applied in any sub-field was $\sim$0$\farcs$9 although in most cases, it was $\sim$0$\farcs$5. The remaining astrometric offset and error are small ($<0\farcs3$ in each optical sub-field) and are shown in Figure~\ref{fig:offset_corr}. In Table~2, we include the corrected X-ray source positions. The optical positions remain unchanged.

Objects brighter than $R\approx$17 are saturated in the typical NDWFS imaging exposures. This can result in large uncertainties in the measured positions and optical magnitudes of these sources. To correct the positions in the small number of affected sources, we matched the X-ray catalog to objects in the 2MASS catalog which has more accurate astrometry than shallow optical surveys such as the DSS. We applied the same astrometric offset that was used for the X-ray to NDWFS reference frames as well as an additional known offset of 0$\farcs$377 east and 0$\farcs$147 north to shift the 2MASS coordinates to that of the NDWFS. If the position of the 2MASS counterpart was within 10$^{\prime\prime}$ of a bright ($R\leq$17) NDWFS counterpart, we used the corrected position of the 2MASS source for the nearest NDWFS source. 

\subsection{Matching Optical Counterparts}
\label{sec:opt_cparts}

To identify the optical counterpart of each X-ray source, one must decide on the matching criteria to be adopted, i.e., at what angular separation can one be reasonably confident that an optical source is the true optical counterpart of an X-ray source. A compromise is required between having a complete sample and minimizing contamination from unrelated optical sources. Identifying the true optical counterpart is made easier by the excellent positional accuracy of \chandra\ which has typical 90\% uncertainties of $\approx$0$\farcs$6 for a bright ($>$ 10 count) on-axis source. However, the uncertainty in the X-ray position is a strong function of both the off-axis distance of the X-ray source from the X-ray pointing center and the strength of the X-ray source. The deep optical catalogs make it more likely that an X-ray source has a true optical counterpart but the high source density at faint optical apparent magnitudes also increases the risk of chance superpositions. 

To properly quantify these issues, we adopted a Bayesian identification scheme that assigns probabilities to optical sources of being the true optical counterpart rather than using a simple matching criteria. The Bayesian scheme is a more sophisticated approach than using a simple matching radius because it can self-consistently include additional information which is contained within the dataset itself. The basis of our method was to evaluate the probability of each of the optical sources surrounding an X-ray source being more probable than a ``background'' source. Our method also supplies a formal estimate of there being no identifiable optical counterpart down to the optical limits of our survey. The Bayesian source identification method is described in Appendix~\ref{appen}. 

\section{Results}

In this Section, we present the results of the Bayesian matching scheme and compare this to a simple matching algorithm. The results from Monte Carlo simulations are reported. We then present the optical catalog. 

\subsection{Match Statistics}

\subsubsection{Fraction of X-ray Point Sources with Optical Counterparts}

Of the 3,213 X-ray sources detected in the Bo\"otes field, we found a likely optical counterpart for 3,140 (98\%). Because all X-ray sources with near-IR identifications also have optical counterparts, we hereafter refer to both the near-IR and optical counterparts in the NDWFS catalog as the ``optical'' counterparts.  Some of the X-ray sources have more than one potential optical counterpart in our data set. For our analysis of the sample properties in the remainder of the paper, we always used the most probable counterpart. For completeness, however, our catalog includes all candidate optical counterparts with more than a 1\% probability of being the matched source.

\begin{figure}
\begin{center}
\setlength{\unitlength}{1mm}
\begin{picture}(60,70)
\put(-20,-15){\special
{psfile="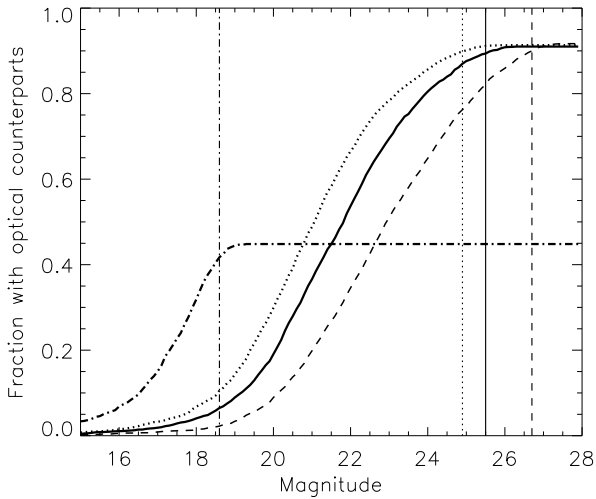" 
vscale=50 hscale=50 angle=0}}
\end{picture}
\end{center}
{\caption[junk]{\label{fig:match_bands2} The fraction of X-ray sources with optical counterparts brighter than a given magnitude in the $B_W$-band (dashed line), $R$-band (solid line), $I$-band (dotted line) and $K$-band (dot-dashed line). The mean 50\% completeness magnitude limits are shown as vertical lines with the same linestyles as above  used for each band. The $K$-band completeness is corrected for its smaller survey area. 
}}
\end{figure}

In Figure~\ref{fig:match_bands2}, we show the fraction of X-ray sources with counterparts brighter than a given magnitude in each optical band. The total fraction of sources with an optical counterpart in each band is less than the total fraction in all bands because some sources are not detected in all the bands. The fraction of $K$-band matches has been corrected for the fact that the area of the $K$-band survey is smaller (only 59\% of the sources that are surveyed in the other bands have coverage in the $K$-band). The number of optical counterparts is similar for each of the $B_W,R$, and $I$ bands. However, only $\sim$45\% of X-ray sources have counterparts in the $K$-band, because the NDWFS $K$-band imaging is significantly shallower than the optical imaging for typical X-ray sources. We discuss the optically non-detected X-ray sources in Section~\ref{sec:nomatch}.

\subsubsection{Comparison with a Simple Matching Algorithm}

We have compared our Bayesian matching scheme with a simple matching scheme in which we define the most likely optical counterpart to be the closest optical match within either 1$\farcs$5 or the positional uncertainty of the X-ray source (Kenter et al.~2005), whichever is greater. Using this simple scheme, we find an optical counterpart for 2875 (89\%) of the X-ray sources. When performing the simple matching scheme on a sample in which we shift the X-ray position by 30$^{\prime\prime}$ (i.e. a ``random'' sample), we find at least one optical counterpart for 38\% of the sources. This high spurious identification rate occurs because the NDWFS images are very deep and have a high number density of faint sources at any position. However, unlike the real data, the optical magnitude distribution in a random sample will be dominated by optically faint sources. This method will therefore overestimate the number of spurious sources. A more rigorous method is presented in Section~\ref{sec:monte}.

The Bayesian matching scheme picks the same optical counterpart as the simple matching scheme in 88\% of cases. Of the 372 X-ray sources whose optical counterparts do not agree, 266 have an optical counterpart in the Bayesian matching scheme but no cataloged optical counterpart in the simple matching scheme; one has an optical counterpart in the simple matching scheme but no cataloged optical counterpart in the Bayesian matching scheme; and 105 have different optical counterparts. If we compare the optical counterpart obtained using the simple scheme to that of either the first or second most likely optical counterpart in the Bayesian matching scheme, only 3\% of the matches disagree. 

We conclude that there is a reasonable agreement between the two matching schemes. The differences are primarily due to the fact that the Bayesian scheme is a more sophisticated algorithm, allowing for the inclusion of more information in the determination of a probable match. For example, the fact that many of the X-ray sources which have no cataloged optical counterpart in the simple scheme have larger positional offsets between the X-ray and optical sources suggests that the positional uncertainty in the simple matching scheme may be underestimated. The X-ray positional error is defined as the 50\% enclosed energy radius divided by the square root of the number of counts (Kenter et al.~2005). We do not expect that the true X-ray position will always fall within the formal positional error and if we increased the matching radius to ensure that we did, we would obtain unacceptable numbers of incorrect matches. In the Bayesian scheme, the positional uncertainty is determined self consistently from the X-ray and optical data themselves. In addition, the simple scheme picks only the nearest cataloged optical counterpart. If there is a brighter optical source (which has a lower probability of being there by chance) slightly further away and the X-ray positional error is large, the Bayesian scheme may assign this to be the true optical counterpart. 

\subsubsection{Monte Carlo Simulations}
\label{sec:monte}

To further test the reliability of our Bayesian technique, we ran Monte Carlo simulations of the optical and X-ray datasets to model the problem of matching the X-ray point sources from the XBo\"otes Survey to the NDWFS optical catalog. The Monte Carlo results provide confidence in our overall approach and a basis for interpreting the results for the real NDWFS/XBo\"otes data. In particular, we use the Monte Carlo tests to understand the completeness of the identifications (hereafter IDs). 

We modeled the X-ray positional uncertainties by a Gaussian (Eqn.~\ref{eqn:psfmod}) with a constant, $\sigma_{a}=0\farcs5$, the PSF width at the pointing center, $\sigma_0=0\farcs5$, and the quadratic growth of the PSF width as one moves off-axis, $\sigma_{600}=4\farcs0$. We produced synthetic background catalogs by taking the XBo\"otes catalog and generating new optical catalogs for the regions around each X-ray source.  We used broken power-law models for the number counts,
\begin{equation}
   {dN \over dm } = A 10^{\alpha(m-m_0)}, 
\end{equation}
empirically normalized to match the observed number counts of the NDWFS field. The total number of sources was normalized to match the $\sim 10^5$ observed sources in the matching regions used for the X-ray sources.  The X-ray sources were modeled with three regions, $\alpha=0.4$ for $15 < m < 20$, $\alpha=0$ for $20 < m < 23$, and $\alpha=-0.3$ for $23 < m < 26$, based on the magnitudes of the most probable optical IDs for the X-ray sources in the central $d_k < 400\farcs0$ regions of the ACIS fields. The background optical source positions were distributed randomly in the matching region for each X-ray source.  To test our ability to recognize X-ray sources genuinely lacking optical counterparts, we assumed that a fraction ($1-f$) of the X-ray sources have no optical counterpart. For the sources with counterparts, the position of the counterpart was randomly drawn based on the ``PSF model'' using the observed X-ray counts and off-axis distance ($C_k$ and $d_k$) for each X-ray source.

\begin{figure}
\centerline{\includegraphics[width=3.0in]{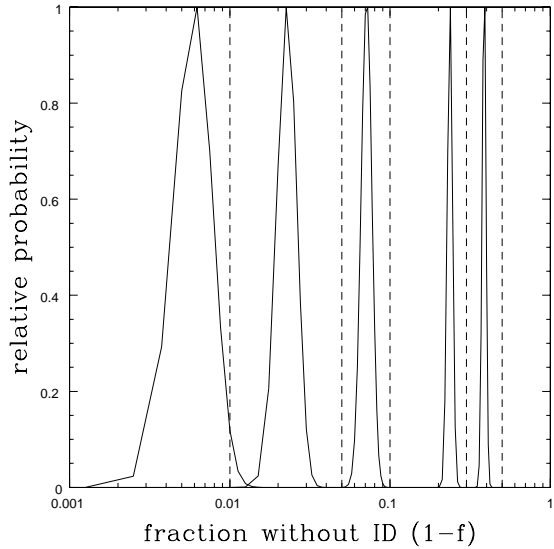}}
\caption{ The posterior probability for the fraction of X-ray sources
  lacking optical counterparts ($1-f$) for Monte Carlo simulations in
  which 99\%, 95\%, 90\%, 70\%, and 50\% of the X-ray sources possess 
  optical counterparts. The vertical dashed lines mark the input 
  values, and in each case the posteriori estimates are shifted 
  by roughly a factor of two towards higher completeness.  If we
  draw the magnitudes of the X-ray sources from the magnitude 
  distribution of the background sources we find no offset.
  These simulations were done with the ``PSF'' parameters fixed to 
  their most likely values.
  }
 \label{fig:sfrac}
\end{figure}

Fig.~\ref{fig:sfrac} shows the results for estimating the fraction of sources with optical counterparts $f$ with the PSF parameters fixed to their most likely values ($\sigma_{a}=\sigma_0=0\farcs5$ and $\sigma_{600}=4\farcs0$).  For the full range of $f=0.99$ to $f=0.5$, the posterior probability distribution for $f$ tracks its true value, but is shifted to higher completeness by approximately a factor of two. The origin of the shift can be traced to the unit ambiguities in the ratio $M_{ik}/B_{k}$ (see Appendix) and the different optical magnitude distributions of the background and X-ray sources. In experiments where we drew the magnitudes of the X-ray sources from the magnitude distribution of the background sources, the posteriori estimates for $f$ were consistent with the input values.  For present purposes, we should regard $f$ as a factor correcting the likelihood balance between ID and no ID for these ambiguities rather than as an absolute measurement of the completeness.  In all experiments, however, the posteriori estimate of $f$ robustly tracked the input completeness.

\begin{figure}
\centerline{\includegraphics[width=3.0in]{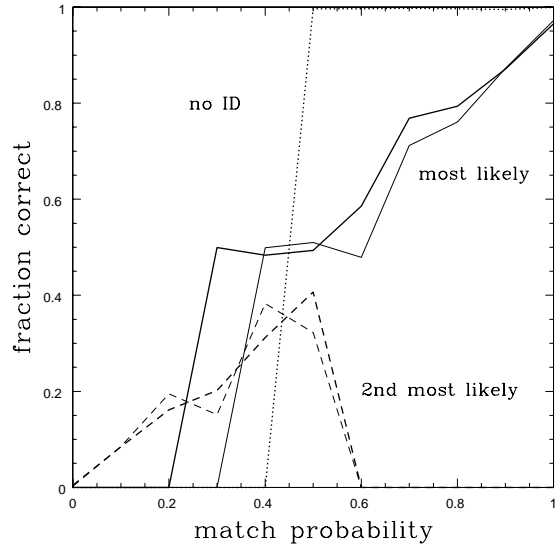}}
\caption{ The solid, dashed and dotted
  curves show the fraction of the time that the most likely, second most likely
  or ``no ID'' optical identification is correct as a function of their probability value for the Monte Carlo simulations with $f=95\%$.
   We show the results for both the full ACIS field (thick line) and the
  inner regions of the ACIS field ($d_k<400\farcs0$; light line).  
  We do not show the ``no ID'' results for the inner regions due to the poor number statistics.    
  }
 \label{fig:correct}
\end{figure}

We now focus on the case with $f=95\%$, as it most closely matches the real data. Fig.~\ref{fig:correct} shows that the probability that an optical identification is correct closely tracks the estimated probability ($P_{ik}$) both for the field as a whole and in the central regions. Over the full field, the true optical source is the most probable match 92\% of the time.  It is the first or second most probable match over 98\% of the time.  As one would expect, most of the matching errors are in the outskirts of the X-ray ACIS pointings where the PSF is large or when the optical counterpart is faint.  For the X-ray sources within 400$^{\prime\prime}$ of the telescope axis, the true optical counterpart is the most probable match over 95\% of the time, and it is the first or second most probable match over 99\% of the time. Similarly, if the most probable optical counterpart has R$<22$~mag, then it is the true counterpart 96\% of the time.  At least in our simulations, misidentified X-ray sources are always matched to another optical source. There was no case of an X-ray source with a counterpart for which having no optical ID was the most probable result.  Of the X-ray sources without optical counterparts, the most probable ID has no optical counterpart roughly 40\% of the time and it is the first or second most probable ID 67\% of the time.  The percentages of correctly matched sources again improve considerably to 48\% and 82\% respectively in the inner parts of the ACIS pointing.  These suffer from significant Poisson uncertainties (because there are only 73 sources without optical IDs) but are typical. Thus, assigning an optical counterpart to X-ray sources genuinely lacking an optical counterpart is a significantly greater problem than the reverse. These results are generic and change little if we consider simulations with larger or smaller incompleteness.

\begin{figure}
\centerline{\includegraphics[width=3.0in]{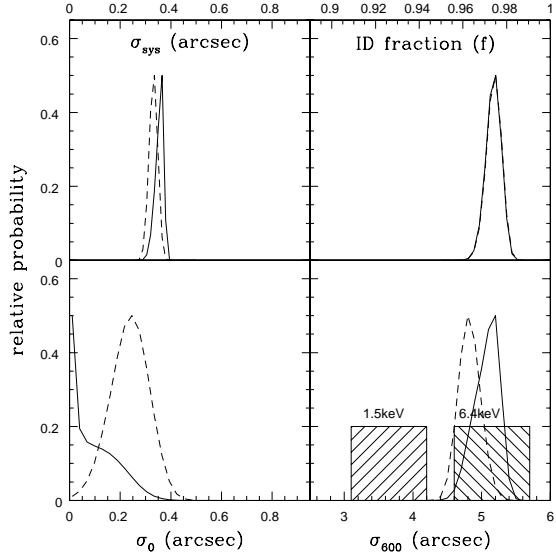}}
\caption{ Posteriori estimates of parameters for the real NDWFS/XBo\"otes data.
  The solid curves use the logarithmic priors and the dashed curves use the 
  CXO priors.  The hatched regions in the $\sigma_{600}$ panel show estimates
  of $\sigma_{600}$ from the 50\% (lower edge) and 90\% (upper edge) enclosed
  energy radii at 1.5 and 6.4~keV (see Eqn.~\ref{eqn:psfmod} for parameter definitions).
  }
 \label{fig:compreal1}
\end{figure}

Fig.~\ref{fig:compreal1} shows the results for the PSF parameters and the completeness.  The completeness estimate is $f=98\%$ independent of whether we use the logarithmic or CXO priors.  Given the bias we observed in the Monte Carlo simulations, this suggests that the real completeness is $f \simeq 95\%$.  The systematic term in the PSF depends little on the priors ($\sigma_0=0\farcs36$ versus $0\farcs33$ for the logarithmic and CXO priors). The other PSF parameters do show a dependence on the priors, with $\sigma_0\simeq0\farcs0$ and $\sigma_{600}=5\farcs2$ for the logarithmic prior and $\sigma_0=0\farcs24$ and $\sigma_{600}=4\farcs8$ with the CXO prior.  
However, despite the apparent parameter differences between the results for the two priors, the posterior probabilities for the identifications are virtually identical.

In summary, the Monte Carlo simulations provide confidence in applying our Bayesian matching scheme to the identification of optical counterparts from the NDWFS to the X-ray sources in the XBo\"otes Survey. By simulating the X-ray sources with true optical counterparts exhibiting similar properties to that expected in the data itself, we demonstrate that the Bayesian method successfully recovers the true optical counterparts as the most probable counterpart 92\% of the time  (and the first or second most probable 98\% of the time). The main errors occur in assigning the wrong optical counterparts or in assigning optical counterparts to sources with no true optical counterpart and this problem is likely to be generic to all matching methods. Although the measured completeness is $\approx$98\%, the true completeness is likely to be f$\approx$95\%.


\subsection{The Optical Catalog and Images of the X-ray Point Sources}

In Table~2 we present a sample of the matched optical and X-ray catalog for the \chandra\ sources in the Bo\"otes field of the NDWFS. The complete version of this table is in the electronic edition of the Journal. The X-ray properties of each X-ray source are presented in Kenter et al~(2005) (the sources can be cross-matched using their X-ray names). The full catalog containing both the X-ray and optical parameters can be found at: ftp://archive.noao.edu/pub/catalogs/xbootes/xbootes$\textunderscore$cat$\textunderscore$xray\\
$\textunderscore$opt$\textunderscore$IR$\textunderscore$21jun$\textunderscore$v1.0.txt and off the NDWFS homepage (http://www.noao.edu/noao/noaodeep/).

The best matching criteria depends on the science to be undertaken and for this reason, we include in this data release all multiply matched sources (with $>$1\% probability of being the correct optical counterpart) as well as information regarding their optical magnitude, distance to the X-ray source and probability of being the optical counterpart. Within our electronic table we include the following columns: \\

\noindent (1) Name of X-ray source\\
(2) X-ray Right Ascension 2000 shifted to NDWFS astrometric frame (hours)\\
(3) X-ray Right Ascension 2000 shifted to NDWFS astrometric frame (minutes)\\
(4) X-ray Right Ascension 2000 shifted to NDWFS astrometric frame (seconds)\\
(5) X-ray Declination 2000 shifted to NDWFS astrometric frame (degrees)\\
(6) X-ray Declination 2000 shifted to NDWFS astrometric frame (minutes)\\
(7) X-ray Declination 2000 shifted to NDWFS astrometric frame (seconds)\\
(8) X-ray Right Ascension 2000 shifted to NDWFS astrometric frame (total degrees)\\
(9) X-ray Declination 2000 shifted to NDWFS astrometric frame (total degrees)\\
(10) Name of X-ray pointing\\
(11) Probability of X-ray source having an optical counterpart\\
(12) Number of optical sources with $>$1\% probability of being associated with the X-ray source\\
(13) Optical rank (1=most probable in running order to least probable)\\
(14) Bayesian probability of match\\
(15) Identification flag (1 -- optical ID; -1 -- no optical ID. In cases of no optical ID being the most probable identification, the following flag values have been applied manually: -3 -- no optical ID, but source obscured by nearby star / missing data; -2 -- no optical ID, but X-ray position is close to optically bright galaxy (source is either obscured by or associated with the galaxy); -1 -- 'true' no optical ID (optical image is truly blank)). \\
(16) NDWFS optical name (Null if no optical ID)\\
(17) Optical (NDWFS) Right Ascension 2000 (hours; Null if no optical ID)\\
(18) Optical (NDWFS) Right Ascension 2000 (minutes; Null if no optical ID)\\
(19) Optical (NDWFS) Right Ascension 2000 (seconds; Null if no optical ID)\\
(20) Optical (NDWFS) Declination 2000 (degrees; Null if no optical ID)\\
(21) Optical (NDWFS) Declination 2000 (minutes; Null if no optical ID)\\
(22) Optical (NDWFS) Declination 2000 (seconds; Null if no optical ID)\\
(23) Optical (NDWFS) Right Ascension 2000 (total degrees; Null if no optical ID)\\
(24) Optical (NDWFS) Declination 2000 (total degrees; Null if no optical ID)\\
(25) Offset of Optical and X-ray positions (Null if no optical ID)\\
(26) $B_W$-band magnitude (Vega; Null: no optical ID, $-$1: no data in this band\footnote{ This may occur for a small number of sources in which the position on the optical image coincides with a masked out region due to a nearby bright star, cosmic ray, or other problem}, 0: no detection in this band (but detected in another band))\\
(27) $R$-band magnitude (Vega; Null if no optical ID, $-$1: no data in this band$^1$, 0: no detection in this band (but detected in another band))\\
(28) $I$-band magnitude (Vega; Null if no optical ID, $-$1: no data in this band$^1$, 0: no detection in this band (but detected in another band))\\
(29) $K$-band magnitude (Vega; Null if no optical ID, $-$1: no data in this band (generally because there is no $K$-band image coverage), 0: no detection in this band)\\
(30) $B_W$-band 50\% completeness limit (-1: no data in this band$^1$)\\
(31) $R$-band 50\% completeness limit (-1: no data in this band$^1$)\\
(32) $I$-band 50\% completeness limit (-1: no data in this band$^1$)\\
(33) $K$-band 50\% completeness limit (-1: no data in this band)\\
(34) Optical Stellarity Parameter (taken from from either the $R$, $B_W$, or $I$-band catalogs in that order of preference; Null if no optical ID)\\
(35) Name of optical sub-field in which source is found\\

\addtocounter{table}{+1}

\begin{figure*}
\begin{center}
\setlength{\unitlength}{1mm}
\begin{picture}(60,250)
\put(-60,-20){\includegraphics{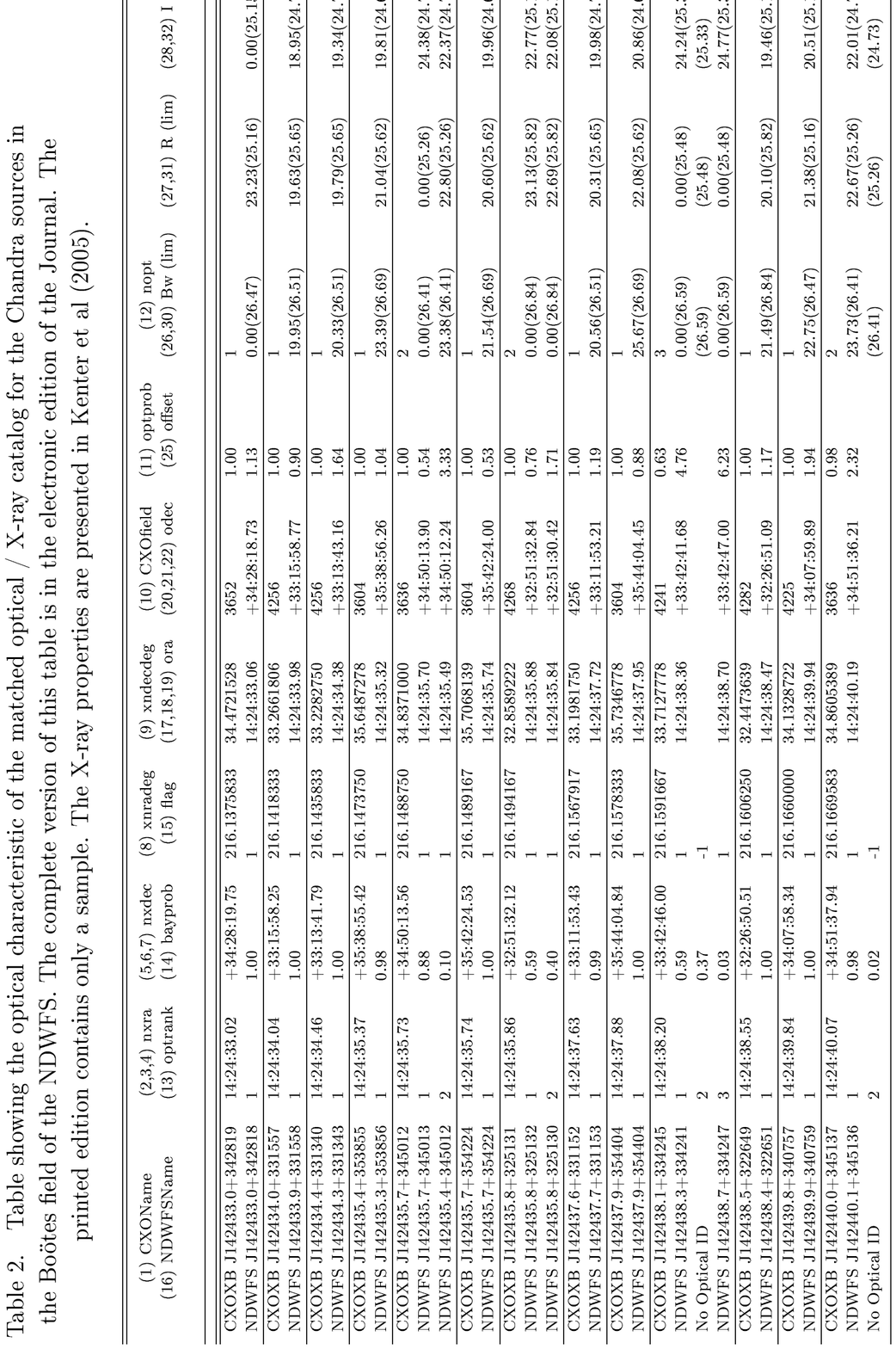}}
\end{picture}
\end{center}
\end{figure*}

\begin{figure*}
\begin{center}
\setlength{\unitlength}{1mm}
\begin{picture}(60,180)
\put(-60,-15){\special
{psfile="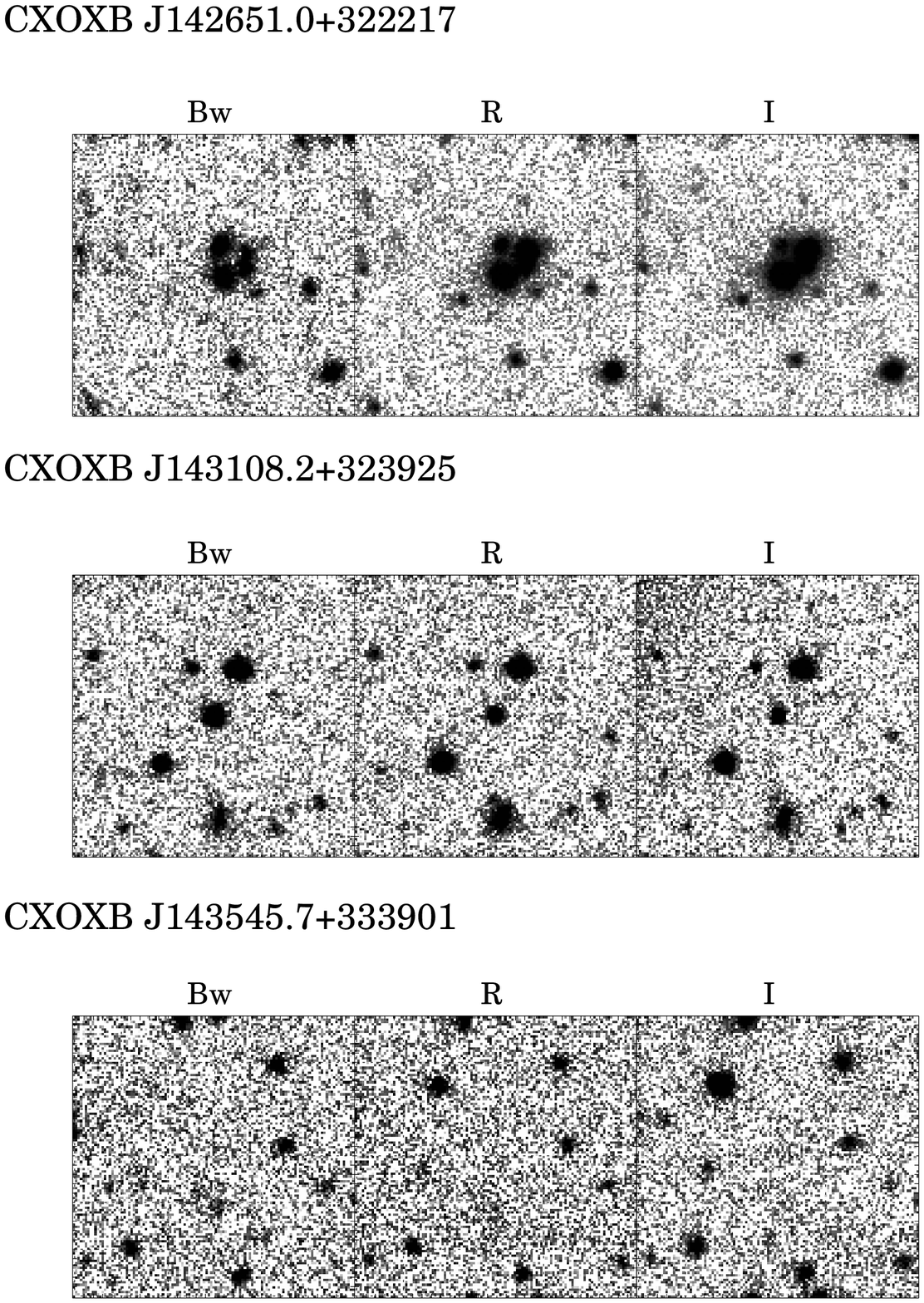" 
vscale=80 hscale=80 angle=0}}
\end{picture}
\end{center}
{\caption[junk]{\label{fig:fcharts} Example optical finding charts for a bright galaxy possibly in an interacting system (top), an optical point source (middle) and an optically non-detected X-ray source (bottom). The finding charts are 30$^{\prime\prime}$ wide and centered on the X-ray position, with North up and East to the left. The X-ray positional uncertainties are 1$\farcs$25, 0$\farcs$86, and 1$\farcs$01 from top to bottom.  
}}
\end{figure*}

Examining the optical images of the X-ray sources is important both in verifying that the matching criteria are sensible and in determining the nature of the X-ray sources. We extracted $B_W$, $R$, $I$,and $K$ cut-out images for all the XBo\"otes X-ray sources with unambiguous matches to check the matches by eye. In Figure~\ref{fig:fcharts}, we show examples for a bright galaxy which appears to be either an interacting galaxy or a galaxy at the center of a cluster, an optical point source and an optically non-detected X-ray source. Optical cut-out images can be obtained using the NOAO cut-out service located at http://archive.noao.edu/ndwfs/cutout-form.html.

\section{The X-ray and Optical Properties of the Matched Sample}

In this Section, we investigate the X-ray and optical properties of the matched population. We use the optical morphology to determine the fraction of sources whose optical light is dominated by point-like emission (AGN and stars) and extended emission (galaxies whose AGN emission is obscured from view). By plotting their color-color distribution, we determine the approximate redshift range and galaxy type or amount of reddening of the AGN. We also investigate the properties of AGN whose optical emission is heavily attenuated (defined by sources with a high  X-ray to optical flux ratio, which includes sources with no optical counterpart). 

\subsection{Optical Morphology}

\begin{figure}
\begin{center}
\setlength{\unitlength}{1mm}
\begin{picture}(60,70)
\put(-20,-20){\special
{psfile="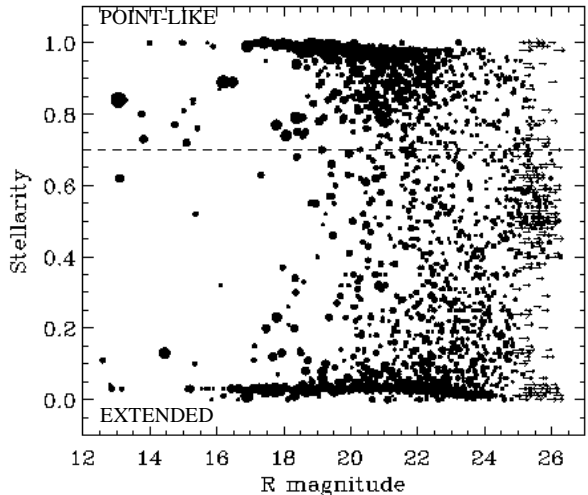" 
vscale=50 hscale=50 angle=0}}
\put(0,5){EXTENDED}
\put(0,58){POINT-LIKE}
\end{picture}
\end{center}
{\caption[junk]{\label{fig:stel} The SourceExtractor stellarity parameter \citep{ber96} plotted against the apparent $R$-band magnitude for all optically matched X-ray sources. The size of both the dots and arrows are scaled relative to the X-ray flux; there is a strong correlation between the optical magnitude and X-ray flux of the sources (see Section~\ref{sec:xopt}). A larger stellarity parameter is obtained for the more point-like objects in the optical image. Arrows denote objects with no $R$-band detection. In these cases we plot the $R$-band 50\% completeness limit for the appropriate optical sub-field; the stellarity parameter has been measured from another band in which a detection is measured. The dotted line shows the boundary that we have assumed in dividing our sample into point sources (stellarity$\ge$0.7) and extended sources (stellarity$<$0.7).
}}
\end{figure}

For our simple morphological analysis, we used the SourceExtractor stellarity parameter \citep{ber96} which has values between 0 (for sources with the most extended radial profiles) and 1 (for point-like sources well characterised by their point spread functions). We made no correction for the variable seeing (0$\farcs$77$\le$FWHM$\le$1$\farcs$53 in the 27 $R$-band sub-fields). In Figure~\ref{fig:stel}, we show the stellarity parameter against the apparent $R$-band magnitude for all optically matched X-ray sources. If we classify all sources with stellarity $\ge$ 0.7 as point-like and all sources with stellarity $<$ 0.7 as (extended) galaxies, there are 1279 optical point sources and 1861 galaxies. SExtractor may fail to correctly classify sources fainter than $R$=23 in the NDWFS data (see, e.g., \citealt{bro03}) and this is apparent in the spread of stellarity values at faint $R$-band magnitudes in Figure~\ref{fig:stel}. If we only consider the sources with $R<$23 counterparts, which comprise 76\% of the sample, there are 1151 point sources and 1304 galaxies.

The point sources are likely to be those in which the optical light is dominated by a central AGN, although the point sources with the highest optical magnitudes are likely to be X-ray bright stars (this is confirmed by optical spectroscopy; Kochanek et al. in prep.). The stellarity of many sources with $R<17$ may be less than 1.0 due to saturation effects. 
The extended sources are those in which the optical light is dominated by the stars in the host galaxy. The vast majority of the sources must contain a powerful AGN due to their luminous X-ray emission and they are likely to be optical type II AGN in which the optical AGN emission is obscured from view. There is a small population of optically bright ($R<17$), X-ray faint galaxies (stellarity $\approx$0); the X-ray emission in these objects is likely to be dominated by high-mass X-ray binaries (HMXBs) whose X-ray emission is correlated with the star formation rate of the galaxy \citep{gri03}. Even in very active star-forming galaxies (SFR$\approx$ 100 M$_{\odot} {\rm yr^{-1}}$), the integrated hard X-ray luminosity from HMXBs should only be L$_x\approx 5 \times 10^{41} {\rm ergs~s^{-1}}$ \citep{gri03}. Because the XBo\"otes survey is relatively X-ray shallow, such X-ray emission should only be detectable within galaxies at $z<$0.15.
Although the stellarity parameter becomes unreliable at faint optical magnitudes, there is a significant number of $R<$23 sources with intermediate values of stellarity. These sources could be bright galaxies with a Seyfert-like central AGN. 

\subsection{Color-Color distributions}

\begin{figure*}
\begin{center}
\setlength{\unitlength}{1mm}
\begin{picture}(100,150)
\put(-40,55){\special
{psfile="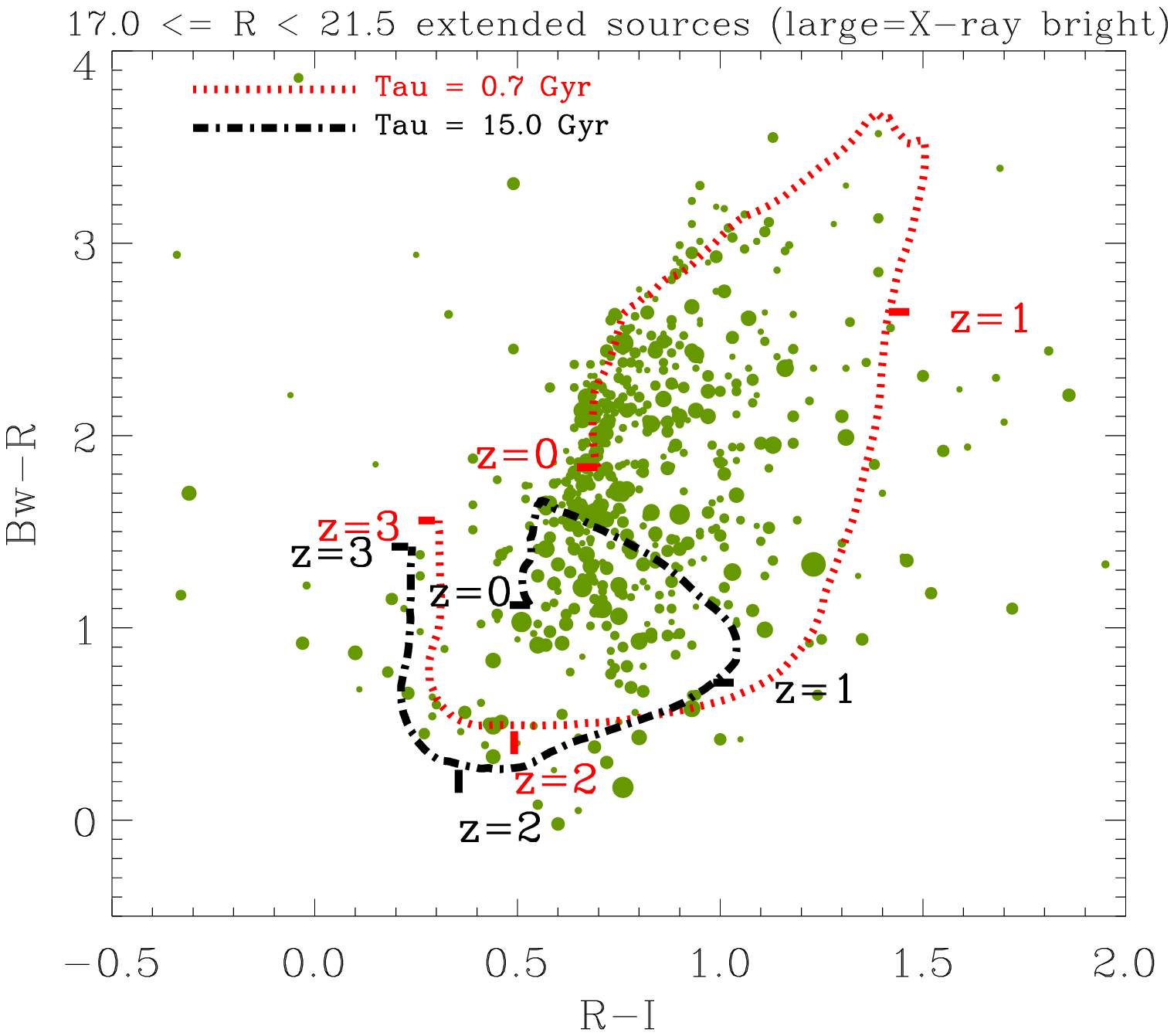" 
vscale=50 hscale=50 angle=0}}
\put(-40,-20){\special
{psfile="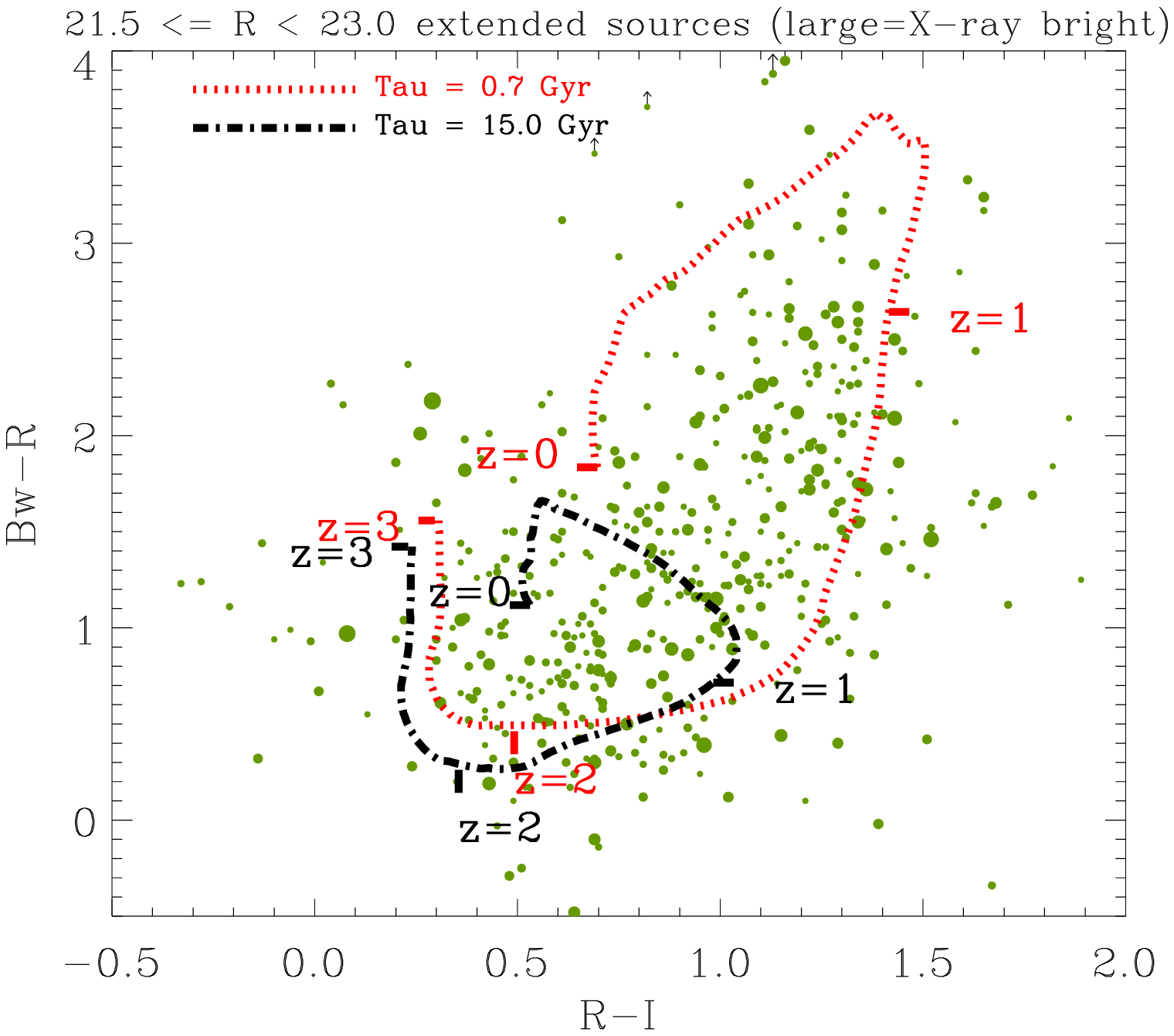" 
vscale=50 hscale=50 angle=0}}
\put(40,55){\special
{psfile="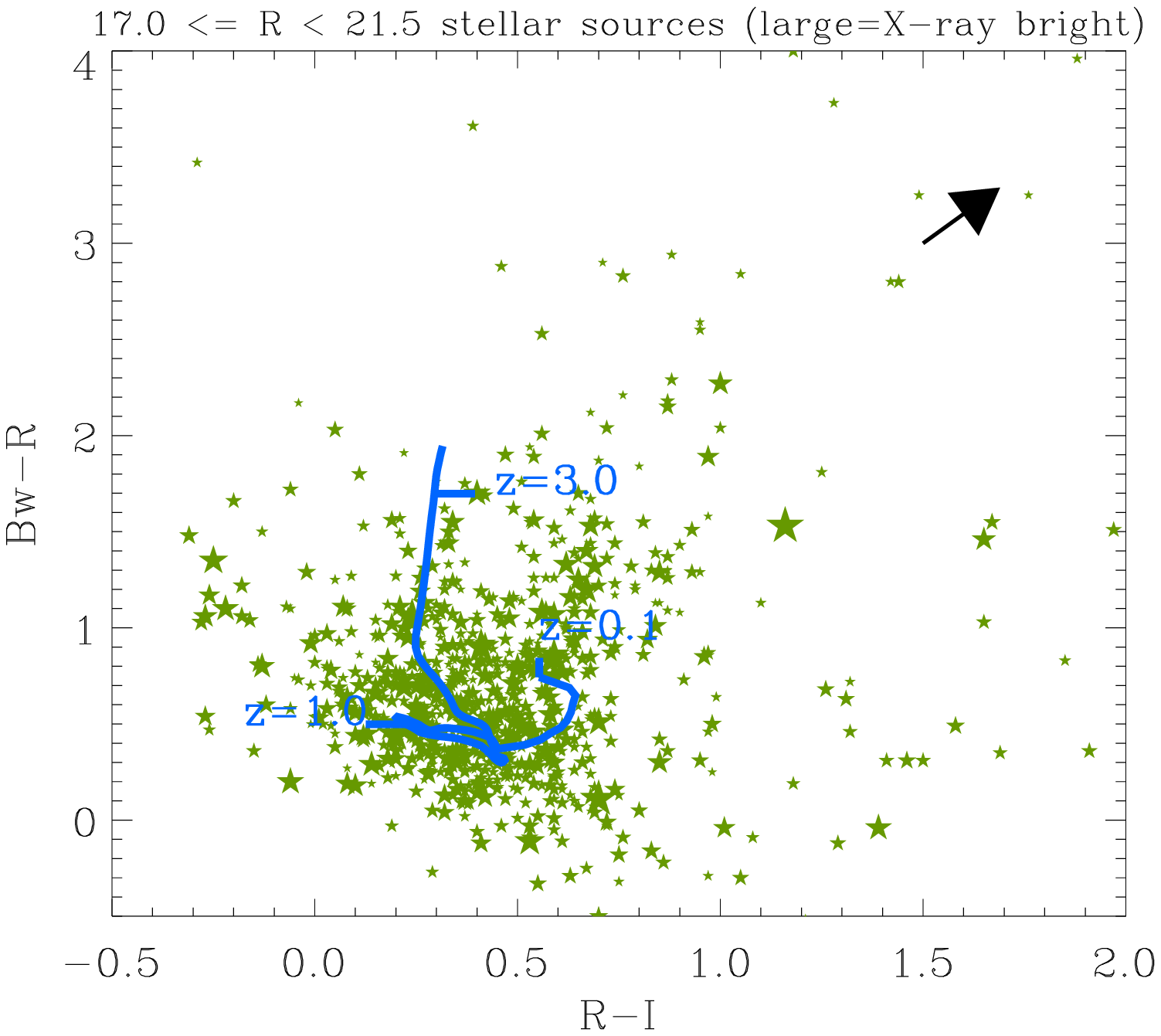" 
vscale=50 hscale=50 angle=0}}
\put(40,-20){\special
{psfile="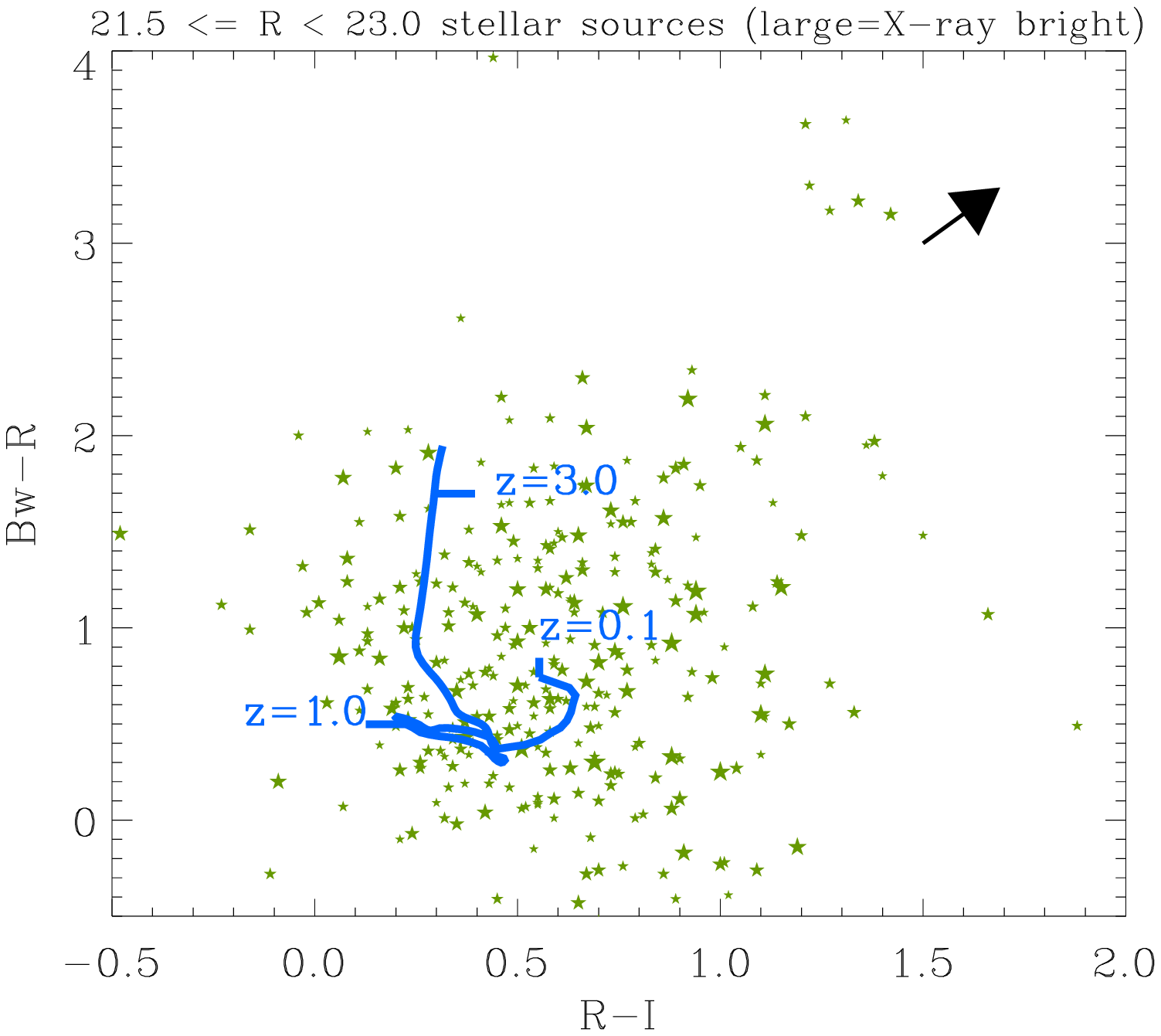" 
vscale=50 hscale=50 angle=0}}
\end{picture}
\end{center}
{\caption[junk]{\label{fig:bmr_rmi_all} Color-color diagrams for all optically matched X-ray sources with an $R$-band detection ($R<$23.0). The sample is split into optically bright ($R<$21.5; top), and optically fainter (21.5$<R<$23.0; bottom) sources. We divide the sample into optically extended sources (stellarity $<$ 0.7; filled circles; left) and optical point sources (stellarity $\ge$ 0.7; stars; right). The symbols are scaled by the logarithm of their X-ray flux (larger symbols denote brighter X-ray sources). Overplotted are the color-color tracks for PEGASE2 model galaxies (left) and the SDSS composite AGN spectrum (right). The arrow denotes the direction of reddening expected if there is additional intrinsic reddening of the sources due to dust.    
}}
\end{figure*}

Color-color diagrams can be used to compare the color distributions of the sample to that expected for model galaxy and AGN templates and hence to infer the general characteristics of a population in terms of spectral type and redshift. In Figure \ref{fig:bmr_rmi_all} we show the $B_W-R$ versus $R-I$ color-color diagrams (Vega magnitudes are used throughout). We divided the sample into optically extended sources (stellarity $<$ 0.7) and optical point sources (stellarity $\ge$ 0.7) and into optically bright (17.0$<R<$21.5) and optically fainter (21.5$<R<$23.0) sources. We do not consider the optically faintest sources because the optical stellarity index is not well determined by our images in these cases. The optically bright sources classified as point sources and as extended sources generally occupy different regions of color-color space. For the extended sources, we overplot the predicted colors of $z$=0$-$3 galaxies using PEGASE2 spectral synthesis models \citep{fio97} that assume exponentially declining star formation rates with $e$-folding times between $\tau$=0.7 and 15 Gyr (``$\tau$'' models) and a formation redshift of $z_f\approx$4. For the point sources (unobscured AGN), we plot the predicted colors of $z$=0.1$-$3.25 quasars by redshifting a composite rest frame SDSS quasar spectrum \citep{van01}. We account for the effects of absorption by intervening QSO absorption systems as a function of redshift using the model of \citet{mad95}.

The majority of the optically bright (17.0$<R<$21.5) extended sources fall on a well defined region of the color-color diagram consistent with them being moderate redshift ($z<1$) galaxies with a range of $\tau$ values (low $\tau$ values are consistent with nearby red early-type galaxies and higher $\tau$ values are consistent with bluer galaxies with higher observed star formation rates). As one would expect, the optically fainter galaxies generally appear to be at higher redshift than the optically brighter galaxies. Their distribution in color-color space is more dispersed. This is probably due to optically faint sources spanning a larger range of redshifts. At fainter optical magnitudes, the uncertainties in the magnitudes become larger and SExtractor starts failing in correctly separating galaxies from point sources and this may also contribute to the larger dispersion in color-color space.

In Figure~\ref{fig:bmr_rmi_all}, we show that the majority of the optically bright (17.0$<R<$23.0) point sources fall in a different region of color-color space than that predicted for stars \citep{pic98}. Comparison to the color-color tracks of a model quasar SED as a function of redshift suggests that the majority of optically bright point sources are quasars at redshifts $z$=0$-$3. At $z>3$, the SED becomes strongly attenuated in the observed $B_W$ band; the color-color track will continue to rise to very red $B_W-R$ colors. There are relatively few sources occupying the color-color space in which one would expect quasars at $z>3$ to lie. This is not surprising given the shallow nature of our X-ray survey compared to the deep \chandra\ pointings and that we are only considering $R<23$ sources. 

Several effects (dust reddening, redshift, relative fractions of galaxy and AGN light, contamination of broad-band colors by strong line emission, mis-classification, etc.) are undoubtedly responsible for the observed scatter in the optical colors of the X-ray sources. Nevertheless, we note the existence of a significant population of red sources in this sample, with some suggestion that the median color of the optically fainter sources is redder than that of the optically brighter sources. For the extended sources, this reddening is probably largely due to the galaxy redshift increasing to fainter magnitudes. For the point sources, however, dust reddening is likely to be the cause of the redder median color.  The most reddened sources may not be present in the optically selected quasar surveys such as the SDSS from which we have used the composite SED to calculate our model tracks. \citet{ric03} and \citet{hop04} find that the reddening of quasars is dominated by SMC-like dust at the quasar redshift. We use the parametric extinction laws within the SMC of \citet{pei92} to determine how this reddening would affect the position of quasars in color-color space. Indeed, it appears that there may be an increasingly large fraction of reddened quasars at fainter optical fluxes. A visual examination of the small number of sources with extreme colors ($B_W-R>3$) reveals that they are genuine sources with very faint $B_W$ emission rather than sources with problems in their optical photometry or a significant contribution from stellar light. This suggests that their UV radiation is strongly attenuated either because they are at high redshift or because they have significant amounts of dust which preferentially absorbs the light at low wavelengths. 

\subsection{The X-ray to Optical Flux Ratio of the matched sample}
\label{sec:xopt}

Traditional optically selected samples of AGN may only detect a fraction of the overall population because in many sources dust obscures the AGN core from view at optical wavelengths. We can study the properties of this optically obscured population based on samples of X-ray selected AGN with a high X-ray to optical flux ratio, defined as: 

\begin{equation}
\log f_x/f_o = \log(f_x)+\frac{R}{2.5}-9.53,
\end{equation}

\begin{figure*}
\begin{center}
\setlength{\unitlength}{1mm}
\begin{picture}(100,70)
\put(-40,-15){\special
{psfile="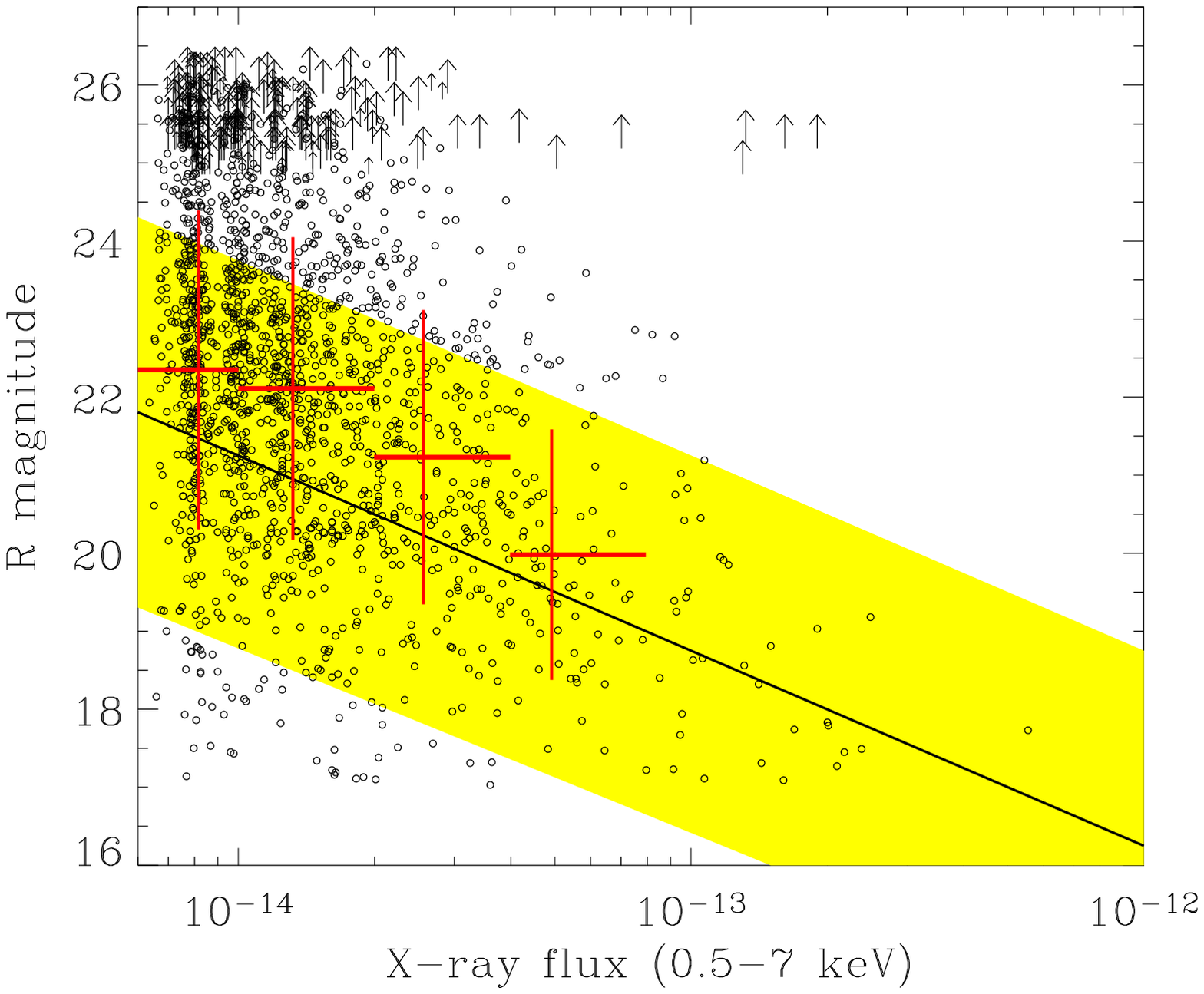" 
vscale=50 hscale=50 angle=0}}
\put(40,-15){\special
{psfile="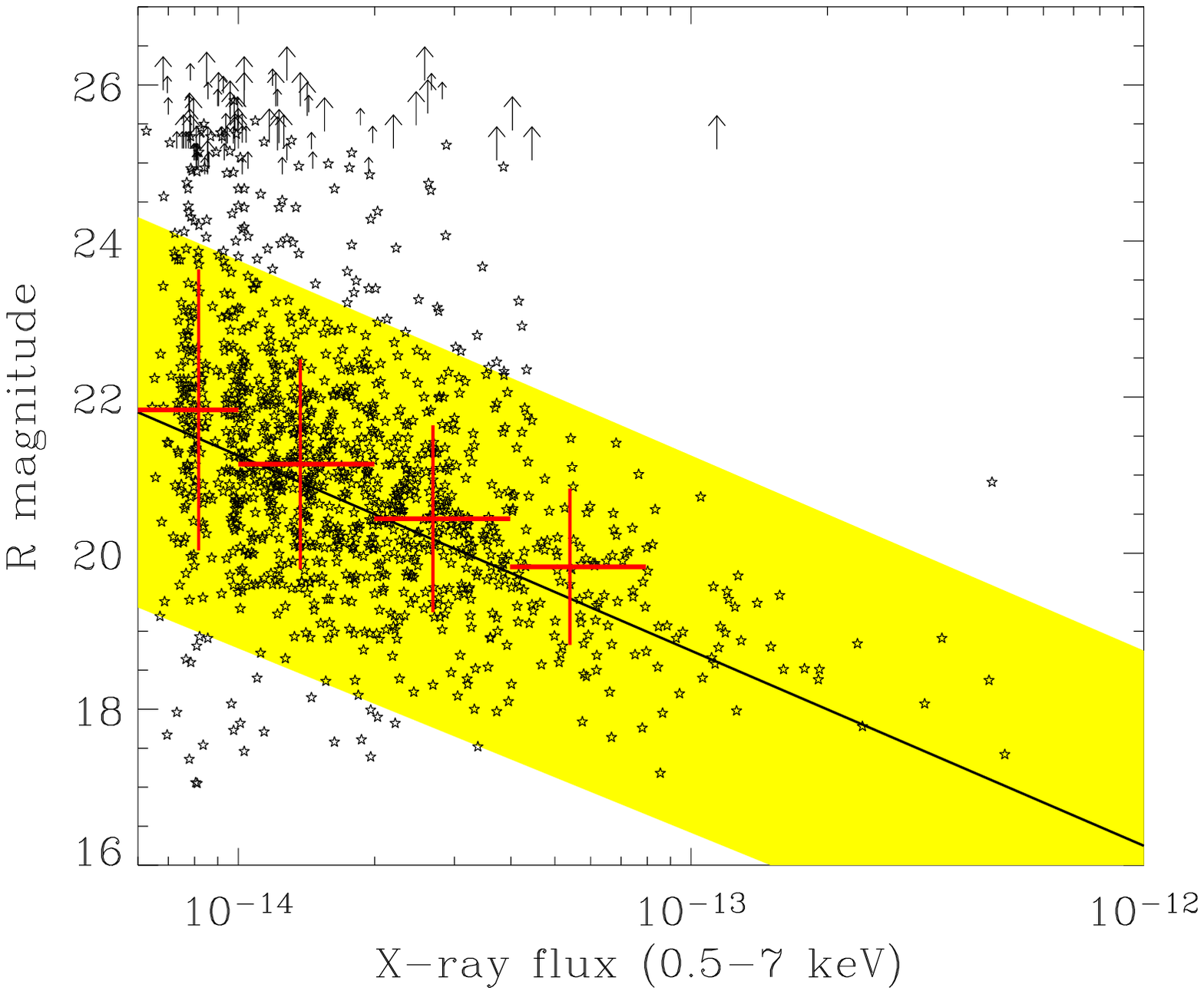" 
vscale=50 hscale=50 angle=0}}
\end{picture}
\end{center}
{\caption[junk]{\label{fig:xoratio} The total (0.5-7 keV) X-ray flux against $R$-band magnitude for all extended sources (stellarity $<$ 0.7; circles; left) and point sources (stellarity $\ge$ 0.7; stars; right). Note that the small amount of vertical structure at low X-ray fluxes is due to quantized source counts. We only plot sources with $R>$17.0 as saturation effects may result in inaccurate optical magnitudes in the brightest cases. Sources with $R<23$ will have less reliable stellarity values. Arrows denote objects with no $R$-band detection. In these cases we plot the $R$-band limit for the appropriate optical sub-field. Small arrows denote objects with a detection in another optical band. Large arrows denote optically non-detected sources. Because there is no information on their stellarity, these sources are plotted on both figures. The (yellow) shaded region denotes the region in which 0.1$<f_x/f_o<$10; our $f_x/f_o>$10 sample lies above this region. The large crosses show the median $R$ magnitude and standard deviation around the mean for different X-ray flux bins.
}}
\end{figure*}

\noindent \citep{mch03} where $f_x$ is the total (0.5-7 keV) X-ray flux in units of $10^{-15} {\rm ergs} ~{\rm cm}^{-2} ~{\rm s}^{-1}$ and $R$ is the apparent $R$-band magnitude. In Figure~\ref{fig:xoratio}, we show the X-ray flux against $R$-band magnitude. Overplotted is the region $0.1<f_x/f_o<10$. Star-forming galaxies tend to fall in the region $f_x/f_o<0.1$ (e.g., \citealt{gia01}; \citealt{hor01}) and optically un-obscured AGN tend to populate the shaded region (0.1$<f_x/f_o<$10) (e.g., \citealt{bar03}; \citealt{ste04}; \citealt{sil04}). The $f_x/f_o>$10 sources are expected to be the optically obscured AGN population (although high redshift X-ray sources may also have large $f_x/f_o$). The optically non-identified sources (Section~\ref{sec:nomatch}) fit within this category. 

\begin{figure}
\begin{center}
\setlength{\unitlength}{1mm}
\begin{picture}(60,70)
\put(-20,-15){\special
{psfile="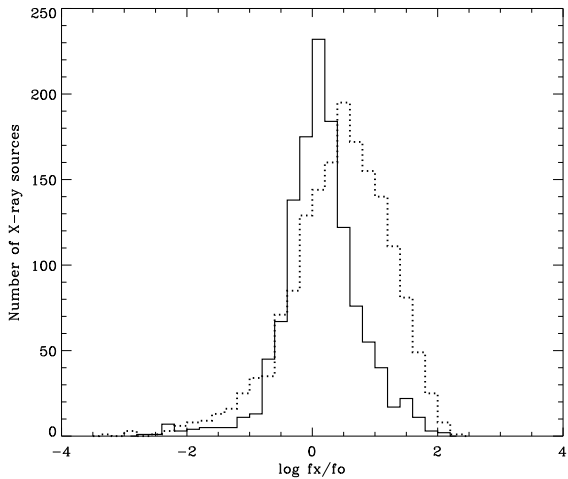" 
vscale=50 hscale=50 angle=0}}
\end{picture}
\end{center}
{\caption[junk]{\label{fig:xohist} Histogram of the number of optically extended sources (obscured AGN; dotted line) and optical point sources (unobscured AGN; solid line) as a function of the X-ray to optical flux ratio for all X-ray sources with a detection in the $R$-band. 
}}
\end{figure}

We calculate the median $R$-band magnitude for different X-ray flux bins (using logarithmic binning). We find that the median value for the optical point sources lies along the $f_x=f_o$ line (see Figure~\ref{fig:xoratio}). The median value for the optically extended sources lies further above the $f_x=f_o$ line, but the scatter is large enough that the median value is still consistent with $f_x=f_o$. In Figure~\ref{fig:xohist}, we illustrate how the extended sources have a larger mean X-ray to optical flux ratio than the point sources. A KS-test \citep{pre92} shows that the X-ray to optical flux ratio distributions of the two populations are drawn from a different underlying distribution at a $>$99.9\% significance. This is probably due to the fact that for the extended sources, the optical emission is from the stellar population (i.e., the host galaxy light dominates either because it is optically brighter or because the central AGN is completely optically obscured) whereas in the case of the point sources, the optical emission is dominated by an optically less obscured AGN. One must be careful because the optical classification may be unreliable at $R>$23 (see Figure~\ref{fig:stel}) and this may affect the classification of sources with $f_x/f_o\ge$5. However, our conclusions still hold when considering only $R<$23 sources (albeit with larger uncertainties).

We define the 773 sources with $f_x/f_o>$10 as our candidates for an optically obscured AGN population (we have omitted the 26 sources that are not truly optically blank; see Section~\ref{sec:nomatch}). The population consists of 510 sources that are detected in $R$, 217 that are only detected in another optical band, and 47 that are optically non-detected in all optical bands to the limits of our survey. All of the non-detected sources have $f_x/f_o>$10. The majority of the $f_x/f_o>$10 sources (643/773) have extended optical emission (stellarity $<$0.7), suggesting that the optical light is dominated by the stars in the host galaxy (i.e. the optical emission from the AGN is heavily obscured). This is also true for the $R<$23 sources (37/47 are optically extended), suggesting that the unreliability of the stellarity parameter at fainter optical magnitudes is not biasing this result. We note that for the sources with low X-ray counts, the X-ray flux may have been over-estimated due to Eddington bias. This effect is discussed in Kenter et al.~(2005) and may mean that $f_x/f_o<10$ for a fraction of our optically obscured AGN sample.  

The X-ray hardness ratio is a useful quantity to approximate the spectral shape of the X-ray emission in cases for which the counting statistics are insufficient to determine the shape of the X-ray spectrum and the redshift distribution is unknown. The X-ray hardness ratio is defined as:
\begin{equation}
HR = \frac{(C_h-C_s)}{(C_h+C_s)},
\end{equation}
\noindent where $C_h$ and $C_s$ are the counts in the hard (2-7 keV) and soft (0.5-2 keV) bands respectively (Kenter at al.~2005). In general, the intrinsic X-ray spectral slope is not thought to vary significantly from source to source. We therefore assume that changes in the hardness ratio can be attributed to a variation in the absorption column of gas which preferentially attenuates the soft X-rays. The hardness ratio spans the range $-$1 (only soft band counts) to 1 (only hard band counts: appropriate for obscured AGN in which the soft photons are absorbed by an intervening column of gas). 

\begin{figure}
\begin{center}
\setlength{\unitlength}{1mm}
\begin{picture}(60,70)
\put(-20,-15){\special
{psfile="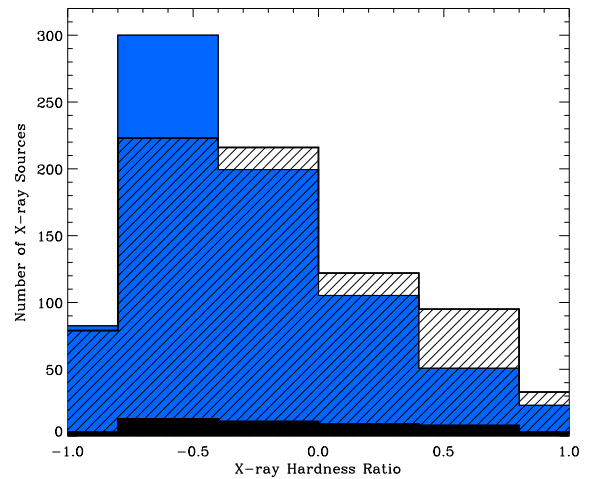" 
vscale=50 hscale=50 angle=0}}
\end{picture}
\end{center}
{\caption[junk]{\label{fig:hrhist} Histogram showing the number of sources as a function of hardness ratio for all X-ray sources with $f_x/f_o>$10 (773 sources; hashed), 0.1$<f_x/f_o<$10 (2280; blue shaded), and the optically unidentified sources (47; black solid). The optically unidentified sources all have $f_x/f_o>$10.
}}
\end{figure}

In Figure~\ref{fig:hrhist}, we show the distribution of X-ray hardness ratio for the optically obscured AGN sample ($f_x/f_o>$10) compared to that of the (renormalized) main AGN sample (0.1$<f_x/f_o<$10). The optically obscured AGN sample exhibits a slightly broader distribution of X-ray hardness ratios whereas the 0.1$<f_x/f_o<$10 sample is more dominated by soft X-ray sources. A KS test rules out the hypothesis that the two hardness ratio distributions are drawn from the same underlying distributions at a probability of $>$99.9\%. This suggests that AGN that are obscured by dust in the optical are also more likely to be obscured in the soft X-ray by gas. However, there is clearly no direct correspondence between these quantities: many of the optically obscured X-ray sources exhibit hardness ratios consistent with them being unobscured in the X-ray and conversely, many optically unobscured X-ray sources exhibit hardness ratios consistent with them being obscured in the X-ray. This is a similar result to that found for a smaller number of sources by \citet{wan04}. The sources with $f_x/f_o>$10 are generally optically fainter (by selection) and may be at a higher redshift than those with 0.1$<f_x/f_o<$10. If this is the case, the negative k-correction in the X-ray will result in similarly obscured sources exhibiting softer X-ray emission, and this may weaken the correlation between sources which appear X-ray obscured and optically obscured. 

\subsection{Sources with no Optical Counterparts}
\label{sec:nomatch}

Of the 3,213 X-ray sources, 73 have the greatest probability of having no cataloged optical counterpart in any of the single-band catalogs ($B_W,R,I,K$) in the NDWFS.  This high completeness is due to the fact that we will correctly optically identify most of our X-ray sources because the XBo\"otes survey is relatively shallow in the X-ray but deep in the optical. We show in Section~\ref{sec:monte} that although the completeness of our matched catalog is $\sim$98\%, the true completeness is likely to be only $\sim$95\%. This is because the number density of faint optical sources is high and so we will have mis-identified some true optically unidentified sources as matches due to chance alignments. We therefore note that our sample of sources with no cataloged optical counterparts is only $\sim$40\% complete and so further optically blank sources exist but are not identified (but this is an unavoidable consequence in any matching technique of such catalogs).

We directly examined the positions of the X-ray sources in the NDWFS images to determine whether these remaining sources truly have no optical counterparts brighter than our survey limits. We divided the sources into those that were near an optically bright star or galaxy (i.e., in regions where the optical imaging or catalog data are compromised or the X-ray emission may come from a ULX in the outskirts of the galaxy) and those that appear to be truly optically blank to the NDWFS survey limits. Ten sources lie close to a bright star or galaxy, but the star or galaxy was not a plausible identification because of the accuracy of the X-ray and optical positions. Because the X-ray position is offset from the optical position of the star or galaxy, the true optical counterpart is probably obscured by the bright star or galaxy, which happens to be near our line of sight to the X-ray source. 
Throughout our analysis, we have assumed that the intrinsic positions of the X-ray and optical emission are identical. However, in some cases, the X-ray emission may be coming from sources offset from the center of the galaxy. Sixteen sources are near to a low-redshift, bright galaxy which is either obscuring the true source from view (as in the above sources), or host to the X-ray source. In the latter case, the X-ray source could be a bright high-mass X-ray binary (HMXB) which is offset from the center of the galaxy. Such sources have been observed in the \chandra\ Deep Field \citep{hor04}. However, the brightest X-ray binaries found in the local Universe have X-ray luminosities no larger than $\sim 10^{40}~{\rm ergs~s^{-1}}$ \citep{gri03} and so for HMXBs to be individually detected above the flux limit of our survey, the galaxy would have to be at low redshift: $z$=0.02 or nearer.

For the remaining 47 of the optically unidentified X-ray sources, there is clearly no optical counterpart in any of the NDWFS bands which could be associated with the X-ray emission. We expect some of these sources will be spurious X-ray detections. From analysis of archival ACIS-I background data, we expect $\approx$35 X-ray sources in the XBo\"otes survey to be spurious (Kenter et al.~2005). Our Monte Carlo simulations suggest that spurious sources will be matched to an optical source approximately 60\% of the time (see Section~\ref{sec:monte}); therefore we expect only $\approx$14 (30\%) of the optically unidentified X-ray sources to be spurious X-ray detections. 

The following properties of the optically unidentified X-ray sources give us further confidence that our sample is not dominated by spurious X-ray detections. We measured the 4$^{\prime\prime}$ diameter aperture magnitude for the 47 optically non-identified X-ray sources and found a significant signal in the majority of cases; 22, 29, and 33 of these sources have $> 3\sigma$ signal over the local background in the $B_W$,$R$, and $I$-band respectively. We expect the majority of the spurious X-ray sources to fall at large off-axis distances and to have X-ray counts close to the survey limit (Kenter et al.~2005). The X-ray sources with no optical identification have a similar distribution in their off-axis distances to the X-ray sources with optical counterparts, suggesting that they are not dominated by spurious detections. In general, they have lower X-ray counts than sources with optical counterparts (with a median X-ray counts of 5 compared to 6 for the sources with optical counterparts) but this may be because they are at higher redshift and/or are obscured in both the optical and soft X-ray band.  

We conclude that these sources are most likely to be dominated by powerful high redshift and/or optically obscured quasars. They are an extreme subset of the $f_x/f_o>$10 sources discussed in Section~\ref{sec:xopt}. Collectively, the optically unidentified sources have a hardness ratio of $-$0.08$\pm$0.005 (135 soft counts, 116 hard counts). A KS-test shows that, like the $f_x/f_o>$10 sources, the hardness ratio distribution is different from that of the 0.1$<f_x/f_o<$10 sources at a $>$99.9\% probability (see Figure~\ref{fig:hrhist}). This suggests that like the $f_x/f_o>$10 sources, they have, on average, large column densities of gas obscuring the soft X-ray emission.

\section{Summary and Discussion}

We have presented the catalog of optical counterparts of the 3,213 X-ray point sources in the Bo\"otes field of the NOAO Deep Wide-Field Survey. Our Bayesian technique finds optical counterparts for 98\% of the X-ray sources. This provides the basis for further investigation of the properties of the X-ray sources. In particular, a large program of optical spectroscopy with the Hectospec instrument on the MMT is underway, targeting all ($\approx$1900) X-ray sources with $R<$21.5 or $I<$21.5 optical counterparts (AGES; Kochanek et al. in prep.). 

Approximately half of the XBo\"otes sources are identified with optical point sources;  the other half have extended optical counterparts dominated by extended galaxy light.  The bright ($R<23.0$) optical point sources fall in a region of optical color-color space consistent with them being quasars at $z\le$3. The fainter point sources appear to be redder in their optical colors, suggesting that they may be more obscured QSOs.  The optical colors of the extended sources suggest that they are a combination of $z<$1 early-type galaxies and bluer star-forming galaxies.
 
The large area, X-ray shallow and optically deep nature of the XBo\"otes survey allows us to identify a large sample of 773 bright X-ray sources with high X-ray to optical flux ratios ($f_x/f_o>10$). We interpret these large X-ray to optical flux ratios as resulting from extinction of the optical light by dust. These X-ray sources are generally harder in their X-ray spectra than the bulk of the X-ray source population ($0.1<f_x/f_o<10$), suggesting that AGN that are obscured by dust in the optical are also more likely to be obscured in the soft X-ray by gas. Such a large sample of bright X-ray sources with extreme properties holds much promise for follow-up multiwavelength spectroscopy and detailed studies of obscured AGN. An extreme subset of the $f_x/f_o>10$ population are 47 X-ray sources which have no optical identification down to $R\sim 25.5$, the average 50\% completeness limit of the NDWFS $R$-band images. 

\acknowledgements{We thank the NDWFS survey team, particularly Glenn Tiede, Jenna Claver, Alyson Ford, Emma Hogan, Lissa Miller, and Erin Ryan for assistance in providing the optical and near-IR data in this work. We also thank the observing and instrument support staffs at Kitt Peak National Observatory. Our research is supported by the National Optical Astronomy Observatory which is operated by the Association of Universities for Research in Astronomy, Inc. (AURA) under a cooperative agreement with the National Science Foundation. Support for this work was also provided by the National Aeronautics and Space Administration through Chandra Award Number GO3-41763 issued by the Chandra X-ray Observatory Center, which is operated by the Smithsonian Astrophysical Observatory for and on behalf of the National Aeronautics and Space Administration under contracts NAS8-39073, NAS8-38248, NAS8-10130, and NAS8-39073.}

\appendix

\section{Bayesian Source Identification}
\label{appen}

In this Section, we outline our Bayesian method (developed by CSK.) for identifying the most probable optical counterpart for each X-ray source in the XBo\"otes survey. This method provides a formal approach to assigning a probability to each optical source in the vicinity of the X-ray source being the true optical counterpart (as well as the probability of there being no cataloged optical counterpart down to the limit of the optical survey). In our implementation, the Bayesian method requires information on the X-ray counts and off-axis distance for each X-ray source, the angular distance of all optical sources from the X-ray source in the vicinity of each X-ray source, and their optical magnitudes. It then uses the data itself to self-consistently estimate each individual X-ray positional error, the probability of an optical source of a given magnitude and distance being there by chance, and ultimately the probability of each neighboring optical source being the true optical counterpart. Although the method described here is applied to the specific problem of determining the most probable optical counterpart for each X-ray source in the XBo\"otes survey, we have tried to keep the discussion as general as possible to allow the method to be extended to similar problems. 

We begin by estimating the likelihood of an optical source in the absence of an X-ray source -- i.e., the probability of the background.  From the overall optical field counts we know the differential number counts of sources, $B=dN/dm$, down to the limit of our optical catalog $m_{lim}$.  In practice, we have used the observed NDWFS counts rather than a theoretical model because it implicitly includes the selection effects of the survey and the catalog.  We now want to compute the probability of finding  $k=1 \cdots n_i$ sources with magnitudes $m_k$ in a region of area $A_i$.  Imagine subdividing the region into infinitesimal bins of angular size $\Delta x$ and magnitude $\Delta m$ (see \citealt{kep99}).  Infinitesimal bins are either empty or contain a single source with an expected number of sources in a bin of $N_\alpha = \Delta x^2 \Delta m B(m)$. Thus, the probability of an empty bin is $\exp(-N_{\alpha})$, the probability of a bin containing an object is $N_\alpha \exp(-N_{\alpha})$, and the log-likelihood of the data is

\begin{equation}
  \ln L = \sum_{empty~bins} (-N_\alpha) + \sum_{full~bins} (-N_\alpha + \ln N_{\alpha}).
\end{equation}   
Notice that the $N_{\alpha}$ terms are summed over all bins, while the $\ln N_{\alpha}$ terms are summed only over bins containing sources, so this becomes

\begin{equation}
 \ln L = \sum_{all~bins} (-N_\alpha) + \sum_{full~bins} (\ln (N_\alpha))
\end{equation}

\begin{equation}
  \ln L = - A_i N(m_{lim}) + \sum_{k=1}^{n_i} \ln B_k 
\end{equation}
where $N(m_{lim})$ is the integrated number counts per unit area and the sum $k=1\cdots n_i$ extends over the $n_i$ optical sources in the field. A constant term that depends on the arbitrary bin widths has been dropped.  We imagine doing this background calculation over a very large region so that the first term is unaffected by the inclusion of a known X-ray source in the region.  Thus in the absence of an X-ray source, the probability of the data ($D_i$) given our background model is

\begin{equation}
   P(D_i) = P_{back} = 
       \exp\left(-A_i N(m_{lim})\right) \Pi_{k=1}^{n_i} B_k
\end{equation}

The statistics of an X-ray source in the field are not governed by $P_{back}$. If we can identify X-ray source $i$ and optical source $k$ with probability $M_{ik}$, then we should drop that source from the calculation of the background probability. The calculation will (briefly) appear to be very complex because we must include all possible discrete identifications and then select between them.  The method we now outline is based on \citet{pre97}. While the initial results for the probabilities are very similar to those found by \citet{der76} or \citet{ben83}, our approach has greater formal clarity and lends itself to some useful extensions.

We include all possible identifications by introducing a binary vector $\vec{v}^{ik}$ for the case in which optical source $k$ ($k=1 \cdots n_i$) is identified with X-ray source $i$, where $n_i$ is the total number of possible identifications in the field around the X-ray source.  Each vector is a binary code whose entries are $v^{ik}_\alpha=0$ for $\alpha\ne k$ and $v^{ik}_\alpha = 1$ for $\alpha=k$ (i.e. the true identification). We allow for the possibility that none of the optical sources are the identification by adding the extra vector $\vec{v}^{i0}$ all of whose entries are zero. We also introduce a probability $f$ for an X-ray source possessing an optical identification. The objective of the calculation is to determine the probabilities of the different possible identifications $P(\vec{v}^{ik})=P_{ik}$ and the mean completeness $f$.  Since our possible identifications are exhaustive, the total probability is simply the sum over all the mutually exclusive possibilities:
\begin{equation}
   P(D_i| \vec{v}^{ik},f) =
      \exp\left(-A_i N(m_{lim})\right) 
   \left[ \left(1-f\right)\Pi_\alpha B_\alpha +
     f \sum_{k=1}^{n_i} 
   \Pi_\alpha M_{i\alpha}^{v^{ik}_\alpha} B_\alpha^{(1-v^{ik}_\alpha)}
   \right].
\end{equation}
This may look overly complex, but the structure of the identification vectors $\vec{v}^{ik}$ allows us to simplify it to


\begin{equation}
   P(D_i| \vec{v}^{ik},f) =
      \Pi_\alpha B_\alpha \exp\left(-A_i N(m_{lim})\right) 
   \left[ \left(1-f\right) + 
     f \sum_{k=1}^{n_i} 
   M_{ik} {\Pi_{\alpha\ne k} B_\alpha \over \Pi_\alpha B_\alpha}
   \right],
\end{equation}

\begin{equation}
   P(D_i| \vec{v}^{ik},f) = P_{back}
     \left[ \left(1-f\right) + f\sum_{k=1}^{n_i} { M_{ik} \over B_k} \right].
\end{equation}
The leading term of $1-f$ is the case of no identification ($k=0$), and the sum covers the possibility of identifying X-ray source $i$ with each of the $k=1\cdots n_i$ optical sources.  Thus, the relative likelihoods of the identifications are determined by the ratios of the likelihoods that an optical source is coincident with the X-ray source compared to the likelihood that it is just there by chance.  Since we intend to use a fixed background model we can simply drop the $P_{back}$ term as it will have no effect on the subsequent results.  It should be retained if the model of the background is going to be optimized as part of the later calculations. The simplest model of the probability of optical source $k$ being associated with X-ray source $i$ is a Gaussian model incorporating the distance of the optical source given the X-ray positional error,  
\begin{equation}
       M_{ik} = { 1 \over 2 \pi \sigma_k^2 } 
     \exp\left(  - { 1 \over 2 } { r_{ik}^2 \over \sigma_k^2}\right),
    \label{eqn:match0}
\end{equation}
where $r_{ik}$ is the distance of the optical source from the X-ray source and $\sigma_k$ is the uncertainty in the relative position, which can be modeled by

\begin{equation}
  \sigma_k^2  = \sigma_{a}^2 + 
     { 1 \over C_k} \left[ 
     \sigma_0 + \sigma_{600} \left( { d_k \over \hbox{600\farcs0} } \right)^2 
     \right]^2. 
    \label{eqn:psfmod}
\end{equation}
The first term, $\sigma_{a}$, models any remaining systematic astrometry problems that do
not depend on the X-ray flux.  The second term models the position- and flux-dependent \chandra\ PSF, with $\sigma_0$ giving the width at the pointing center ($d_k=0$) and $\sigma_{600}$ describing the quadratic growth of the PSF width off the axis.  The accuracy with which an X-ray source is centroided improves as the number of counts increases. We therefore scale with the number of X-ray counts $C_k$.  We will refer to this as the ``PSF model''. 

The derivation for multiple ID possibilities is more elaborate and we used Cartesian rather than polar coordinates, but our basic expressions are identical to those of \citet{ben83}.   The probabilities always depend on the ratio $M_{ik}/B_k$, which ought to be dimensionless.  If we use the number counts as a function of magnitude $B_k$ (units mag$^{-1}$~deg$^{-2}$) and the probability for the source position in Eqn.~\ref{eqn:match0} (units deg$^{-2}$), then the ratio is not in some senses dimensionless. This does not matter if we force all sources to have optical IDs ($f\rightarrow 1$), but the probability of finding no ID for a given source and the posterior probability distribution $P(f|D)$ are affected by this problem.  We could include the probability that an X-ray source of a given flux has a given optical magnitude, but this adds more parameters than we presently want to explore. Instead we use 
\begin{equation}
       M_{ik} = { 1\over \sqrt{2\pi}\sigma_m } { 1 \over 2 \pi \sigma_k^2 } 
     \exp\left(  - { 1 \over 2 } { r_{ik}^2 \over \sigma_k^2}\right)
    \label{eqn:match1}
\end{equation}
where $\sigma_m$ is the dispersion of the optical catalog in magnitudes. This simple modification lead to better quantitative results.

\def\bfp{{\bf p}}
So far nothing in our calculation is very explicitly Bayesian.  However, we now want to invert the probability of the data given the model parameters $P(D|\bfp)$, to obtain the probability of the parameters given the data $P(\bfp|D)$.  Doing this inversion requires a prior probability $P(\bfp)$ for the parameters, since probability calculus says that
\begin{equation}
   P(\bfp|D) \propto P(D|\bfp) P(\bfp).
\end{equation}
In all subsequent equations (with a $\propto$ sign) there is an implicit normalization such that
\begin{equation}
 \int d\bfp P(D|\bfp) P(\bfp)=1,
\end{equation} 
which we do not include in the equations because it is too cumbersome. It is the introduction of the priors (the $P(\bfp)$) which tends to generate nervousness about Bayesian calculations.  In our present calculation, if the IDs depend on the priors then one has a much more basic, practical problem than issues of statistical principle.

If we understand our matching model (the $M_{ik}$) perfectly, then the probability of the source IDs given the data is
\begin{equation}
   P(\vec{v}^{ik},f|D) \propto P(\vec{v}^{ik}) P(f) P(D_i| \vec{v}^{ik},f). 
\end{equation}
We assume that the prior likelihoods of all possible identifications are equal ($P(\vec{v}^{ik})=1$). For the probability of an X-ray source having an optical identification ($P(f)$), we use both logarithmic priors and rough Gaussian priors based on the known characteristics of the \chandra\ point spread function ($\sigma_0$=0$\farcs$6$\pm$0$\farcs$2 and $\sigma_{600}$=5$\farcs$0$\pm$1$\farcs$6). In practice, the use of these different priors yield similar results. To start with, we assume that the average probability $f$ of an X-ray source having an identification is constant. The probability of identifying X-ray source $i$ optical source $k$ is then
\begin{equation}
    P_{ik} = P(\vec{v}^{ik}|D) = f { M_{ik} \over B_k } 
     \left[ \left(1-f\right) + f \sum_{l=1}^{n_i} { M_{il} \over B_l} \right]^{-1} 
   \label{eqn:simp1}
\end{equation}
while the probability of having no ID is
\begin{equation}
    P_{i0} = P(\vec{v}^{i0}|D) = \left(1-f\right)
     \left[ \left(1-f\right) + f \sum_{l=1}^{n_i} { M_{il} \over B_l} \right]^{-1}
   \label{eqn:simp2}
\end{equation}
and these probabilities can be evaluated for each individual source. Note that the probability for individual sources is not governed solely by the average probability -- if $M_{il}/B_l$ is a large number, meaning that it is extremely unlikely to find an optical source that closely associated with the X-ray source by chance, then $P_{ik} \rightarrow 1$ and $P_{i0} \rightarrow 0$ independent of $f$.  But $f$ should really just be treated as another Bayesian parameter, so we should estimate our identification probabilities by marginalizing over $f$ rather than using a fixed value.  Similarly, by marginalizing over all possible identifications, we obtain an estimate of the probability distribution for $f$. We can also include additional, global parameters.  For example, if there are uncertainties in the accuracy of the astrometry, then the matching probability $M_{il}$ may also have some unknown or ill-constrained parameters.    

Once we include a set of global parameters, which we will call $\bfp$, then we must work with the probability of matching all the sources rather than that of matching individual sources,
\begin{equation}
    P(\vec{v}^{il},\bfp|D) \propto P(\bfp) \Pi_i P(D_i| \vec{v}^{il},\bfp),
\end{equation}
where we now take the product $i=1\cdots N$ over all the X-ray sources and add a prior $P(\bfp)$ for the new parameters.  The expression assumes a uniform prior for the identification vectors $P(\vec{v}^{il})=1$.\footnote{This may not be appropriate in all circumstances. For example, if spectroscopy has shown that some of the candidate IDs for the X-ray sources are quasars, then the priors for that field should reflect the quasars are almost certainly the proper identifications.} In order to estimate the probability of identifying optical counterpart $l$ with X-ray source $i$, we must now marginalize the probability over all the parameters $\bfp$ as well as all the other X-ray sources and their identifications. If we define     
\begin{equation}
    P_{il} = P(\vec{v}^{il}|D) \propto p_{il} = \int d\bfp P(\bfp) 
     f { M_{il} \over B_l } 
         \Pi_{m\ne i} P(D_i| \vec{v}^{ik},\bfp), 
\end{equation}
for optical source $l$, and
\begin{equation}
    P_{i0} = P(\vec{v}^{il}|D) \propto p_{i0} = \int d\bfp P(\bfp) 
         \left( 1-f \right) \Pi_{m\ne i} P(D_i| \vec{v}^{ik},\bfp), 
\end{equation}
for the possibility of no identification, then the probability of identifying optical source $l$ with X-ray source $i$ is 
\begin{equation}
      P_{il} = { p_{il} \over p_{i0} + \sum_{l=1}^{n_i} p_{il} }
\end{equation}
and the probability of having no identification is
\begin{equation}
      P_{i0} = { p_{i0} \over p_{i0} + \sum_{l=1}^{n_i} p_{il} }
\end{equation}
where we have now included the normalizing denominators.  If the mean probability of an identification $f$ is the only parameter, the only difference from Eqns.~\ref{eqn:simp1} and \ref{eqn:simp2} is that the identification probabilities for X-ray source $i$ are weighted by the probability distribution for $f$ estimated from its prior and the data for all other sources rather than using a specified value. Similarly we can determine the probability distributions for the extra parameters by marginalizing over all the other parameters and all the source identifications to find that the probability distribution for parameter $p_m$ is
\begin{equation}
    P(p_m|D) \propto \int d\bfp_{l\ne m} P(\bfp) \Pi_i P(D_i| \vec{v}^{ik},\bfp).
\end{equation}
While some care must be exercised in ordering the calculation and in avoiding variable overflow problems, this algorithm is easily and compactly implemented.  The only significant limitation is that each new parameter $\bfp$ requires an additional integral for the identification probabilities, eventually leading to the usual difficulties accompanying the numerical calculation of many-dimensional integrals.

\bibliography{ms.bbl}

\begin{thebibliography}{6}
\expandafter\ifx\csname natexlab\endcsname\relax\def\natexlab#1{#1}\fi

\bibitem[Alexander et al.(2003)]{ale03} Alexander, D.~M., et 
al.\ 2003, \aj, 126, 539 

\bibitem[Baldi et al.(2002)]{bal02} Baldi, A., Molendi, S., 
Comastri, A., Fiore, F., Matt, G., \& Vignali, C.\ 2002, \apj, 564, 190 

\bibitem[Barger et al.(2003)]{bar03} Barger, A.~J., et al.\ 
2003, \aj, 126, 632

\bibitem[Bauer et al.(2004)]{bau04} Bauer, F.~E., Alexander, 
D.~M., Brandt, W.~N., Schneider, D.~P., Treister, E., Hornschemeier, A.~E., 
\& Garmire, G.~P.\ 2004, \aj, 128, 2048

\bibitem[Becker, White, \& Helfand(1995)]{bec95} Becker, 
R.~H., White, R.~L., \& Helfand, D.~J.\ 1995, \apj, 450, 559

\bibitem[Benn(1983)]{ben83} Benn, C.~R.\ 1983, The 
Observatory, 103, 150 

\bibitem[{{Bertin} \& {Arnouts}(1996)}]{ber96}
{Bertin}, E., \& {Arnouts}, S. 1996, \aaps, 117, 393

\bibitem[Brand et al.(2005)]{bra05} Brand, K., et al.\ 2005, 
\apj, 626, 723

\bibitem[Brandt et al.(2001)]{bra01} Brandt, W.~N., et al.\ 
2001, \aj, 122, 2810 

\bibitem[{{Brown} {et~al.}(2003){Brown}, {Dey}, {Jannuzi}, {Lauer}, {Tiede}, \&
  {Mikles}}]{bro03}
{Brown}, M.~J.~I., {Dey}, A., {Jannuzi}, B.~T., {Lauer}, T.~R., {Tiede}, G.~P.,
  \& {Mikles}, V.~J. 2003, \apj, 597, 225

\bibitem[de Ruiter, Willis, \& Arp(1976)]{der76} de Ruiter, H.R., Willis, A.G., \& Arp, H.C., 1976, 
  A\&AS, 28, 211

\bibitem[de Vries et al.(2002)]{dev02} de Vries, W.~H., 
Morganti, R., R{\" o}ttgering, H.~J.~A., Vermeulen, R., van Breugel, W., 
Rengelink, R., \& Jarvis, M.~J.\ 2002, \aj, 123, 1784 

\bibitem[Elston et al.(2005)]{els05} Elston, R., et al.\ 2005, \apj, submitted

\bibitem[Eisenhardt et al.(2004)]{eis04} Eisenhardt, P.~R., 
et al.\ 2004, \apjs, 154, 48

\bibitem[Fioc \& Rocca-Volmerange(1997)]{fio97} Fioc, M., \& 
Rocca-Volmerange, B.\ 1997, \aap, 326, 950 

\bibitem[Freeman et al.(2002)]{fre02} Freeman, P.~E., Kashyap, V., Rosner, R., \& Lamb, D.~Q.\ 2002, \apjs, 138, 185 


\bibitem[Garmire et al.(2003)]{gar03} Garmire, G.~P., Bautz, 
M.~W., Ford, P.~G., Nousek, J.~A., \& Ricker, G.~R.\ 2003, \procspie, 4851, 
28 

\bibitem[Giacconi et al.(2001)]{gia01} Giacconi, R., et al.\ 
2001, \apj, 551, 624

\bibitem[Giacconi et al.(2002)]{gia02} Giacconi, R., et al.\ 
2002, \apjs, 139, 369


\bibitem[Georgakakis et al.(2003)]{geo03} Georgakakis, A., 
Georgantopoulos, I., Stewart, G.~C., Shanks, T., \& Boyle, B.~J.\ 2003, 
\mnras, 344, 161

\bibitem[Green et al.(2004)]{gre04} Green, P.~J., et al.\ 
2004, \apjs, 150, 43

\bibitem[Grimm et al.(2003)]{gri03} Grimm, H.-J., Gilfanov, 
M., \& Sunyaev, R.\ 2003, \mnras, 339, 793

\bibitem[Hasinger et al.(2001)]{has01} Hasinger, G., et al.\ 
2001, \aap, 365, L45

\bibitem[Hoopes(2004)]{hoo04} Hoopes, C.~G.\ 2004, American 
Astronomical Society Meeting, 204,  

\bibitem[Hopkins et al.(2004)]{hop04} Hopkins, P.~F., et al.\ 
2004, \aj, 128, 1112 

\bibitem[Hornschemeier et al.(2001)]{hor01} Hornschemeier, 
A.~E., et al.\ 2001, \apj, 554, 742 

\bibitem[Hornschemeier et al.(2004)]{hor04} Hornschemeier, 
A.~E., et al.\ 2004, \apjl, 600, L147 

\bibitem[{{Jannuzi} \& {Dey}(1999)}]{jan99}
{Jannuzi}, B.~T., \& {Dey}, A. 1999, in "Photometric Redshifts and the
  Detection of High Redshift Galaxies", ASP Conference Series, Vol. 191, Edited
  by R. Weymann, L. Storrie-Lombardi, M. Sawicki, and R. Brunner., 111

\bibitem[Jansen et al.(2001)]{jan01} Jansen, F., et al.\ 
2001, \aap, 365, L1 

\bibitem[Kenter et al.(2005)]{ken05} Kenter, A., et al.\ 
2005, \apjs, 161,1 

\bibitem[Kepner et al.(1999)]{kep99} Kepner, J., Fan, X., Bahcall, N., Gunn, J., Lupton, R., \& Xu, G., 1999, \apj, 517, 78

\bibitem[Kim et al.(2004)]{kim04} Kim, D.-W., et al.\ 2004, 
\apjs, 150, 19

\bibitem[Lehmann et al.(2001)]{leh01} Lehmann, I., et al.\ 
2001, \aap, 371, 833

\bibitem[Lehmer et al.(2005)]{leh05} Lehmer, B.~D., et al.\ 
2005 \apjs, accepted, astro-ph/0506607

\bibitem[Madau(1995)]{mad95} Madau, P.\ 1995, \apj, 441, 18

\bibitem[{{Maiolino} \& {Rieke}(1995)}]{mai95} Maiolino, R.~\& 
Rieke, G.~H., 1995, \apj, 454, 95 

\bibitem[McHardy et al.(2003)]{mch03} McHardy, I.~M., et al.\ 
2003, \mnras, 342, 802

\bibitem[Monet et al.(1998)]{mon98} Monet, D.~B.~A., et al.\ 
1998, VizieR Online Data Catalog, 1252, 0 

\bibitem[{{Moretti} {et~el.}(2003)}]{mor03} Moretti, A., Campana, S., Lazzati, 
D., \& Tagliaferri, G.\ 2003, \apj, 588, 696

\bibitem[Murray et al. (2005)]{mur05} Murray, S.~S., et al. \ 2005, \apj, accepted 

\bibitem[Nandra et al.(2005)]{nan05} Nandra, K., et al.\ 
2005, \mnras, 356, 568


\bibitem[Pei(1992)]{pei92} Pei, Y.~C.\ 1992, \apj, 395, 130 

\bibitem[Pickles(1998)]{pic98} Pickles, A.~J.\ 1998, \pasp, 
110, 863 

\bibitem[Press et al.(1992)]{pre92} Press, W.~H., Teukolsky, 
S.~A., Vetterling, W.~T., \& Flannery, B.~P.\ 1992, Cambridge: University 
Press, 2nd ed.  

\bibitem[Press(1997)]{pre97} Press, W.H.,1997, in Unsolved Problems in Astrophysics, J.N. Bahcall \& J.P. Ostriker, eds., (Princeton: Princeton Univ. Press) 49 
\bibitem[Richards et al.(2003)]{ric03} Richards, G.~T., et 
al.\ 2003, \aj, 126, 1131

\bibitem[{{Risaliti} \& {Elvis} (2004)}]{ris04} {Risaliti}, G. ~\& {Elvis}, M. ,2004, Review for "Supermassive Black Holes in the Distant Universe", Ed. A. J. Barger, Kluwer Academic Publishers

\bibitem[Rosati et al.(2002)]{ros02} Rosati, P., et al.\ 
2002, \apj, 566, 667

\bibitem[Schmidt \& Green(2003)]{sch03} Schmidt, D., \& 
Green, P.\ 2003, ASP Conf.~Ser.~295: Astronomical Data Analysis Software 
and Systems XII, 295, 81 

\bibitem[Silverman et al.(2004)]{sil04} Silverman, J.~D., et al.\ 
2004, \apj, 618, 123

\bibitem[{{Soifer} {et~al.}(2004)}]{soi04} Soifer, 
B.~T.~\& Spitzer/NOAO Team 2004, American Astronomical Society Meeting, 
204

\bibitem[{{Stark} {et~al.}(1992){Stark}, {Gammie}, {Wilson}, {Bally}, {Linke},
  {Heiles}, \& {Hurwitz}}]{sta92}
{Stark}, A.~A., {Gammie}, C.~F., {Wilson}, R.~W., {Bally}, J., {Linke}, R.~A.,
  {Heiles}, C., \& {Hurwitz}, M. 1992, \apjs, 79, 77

\bibitem[Steffen et al.(2004)]{ste04} Steffen, A.~T., Barger, 
A.~J., Capak, P., Cowie, L.~L., Mushotzky, R.~F., \& Yang, Y.\ 2004, \aj, 
128, 1483

\bibitem[Stern et al.(2002)]{ste02} Stern, D., et al.\ 2002, 
\aj, 123, 2223

\bibitem[Stern et al.(2005)]{ste05} Stern, D., et al.\ 2005, 
\apj, 631, 163

\bibitem[Vanden Berk et al.(2001)]{van01} Vanden Berk, D.~E., 
et al.\ 2001, \aj, 122, 549

\bibitem[van Speybroeck et al.(1997)]{van97} van Speybroeck, 
L.~P., Jerius, D., Edgar, R.~J., Gaetz, T.~J., Zhao, P., \& Reid, P.~B.\ 
1997, \procspie, 3113, 89

\bibitem[Wang et al.(2004)]{wan04} Wang, J.~X., et al.\ 2004, 
\aj, 127, 213 


\bibitem[Weisskopf et al.(2002)]{wei02} Weisskopf, M.~C., 
Brinkman, B., Canizares, C., Garmire, G., Murray, S., \& Van Speybroeck, 
L.~P.\ 2002, \pasp, 114, 1 

\bibitem[Wilson et al.(2004)]{wil04} Wilson, G., et al.\ 
2004, \apjs, 154, 107

\bibitem[Worsley et al.(2004)]{wor04} Worsley, M.~A., Fabian, 
A.~C., Barcons, X., Mateos, S., Hasinger, G., \& Brunner, H.\ 2004, \mnras, 
352, L28

\bibitem[Yang et al.(2004)]{yan04} Yang, Y., Mushotzky, 
R.~F., Steffen, A.~T., Barger, A.~J., \& Cowie, L.~L.\ 2004, \aj, 128, 1501

\end{thebibliography}

\end{document}